\documentclass[aps,pre,showpacs,onecolumn,superscriptaddress,amssymb]{revtex4}
\usepackage{graphicx}
\usepackage{makeidx}
\usepackage{amsmath}
\usepackage{comment}
\usepackage{color}
\usepackage{braket}
\usepackage{amssymb}
\usepackage{ulem}
\usepackage{xcolor}
\usepackage{txfonts}

\def\Tr{\mathrm{Tr}}

\newcommand \be{\begin{eqnarray}}
\newcommand \ee{\end{eqnarray}}
\newcommand{\del}{\partial}

\def\dd{\mathrm{d}}

\def\simge{\mathrel{
    \rlap{\raise 0.511ex \hbox{$>$}}{\lower 0.511ex \hbox{$\sim$}}}}
\def\simle{\mathrel{
    \rlap{\raise 0.511ex \hbox{$<$}}{\lower 0.511ex \hbox{$\sim$}}}}

\newcommand*{\colorboxed}{}
\def\colorboxed#1#{%
  \colorboxedAux{#1}%
}
\newcommand*{\colorboxedAux}[3]{%
  \begingroup
    \colorlet{cb@saved}{.}%
    \color#1{#2}%
    \boxed{%
      \color{cb@saved}%
      #3%
    }%
  \endgroup
}
\begin{document}
%%%%%%%%%%%%%%%%%%%%%%%%%%%%%%%%%%%%%%%%%%

\title{Full Dysonian dynamics of the complex Ginibre ensemble}

\author{Jacek Grela} 
\email{jacek.grela@lptms.u-psud.fr} 
\affiliation{LPTMS, CNRS, Univ. Paris-Sud, Universit\'e Paris-Saclay, 91405 Orsay, France}

\author{Piotr Warcho\l{}} 
\email{piotr.warchol@uj.edu.pl} 
\affiliation{M. Smoluchowski Institute of Physics,  Jagiellonian University, PL--30--348 Cracow, Poland}

\date{\today}

%%%%%%%%%%%%%%%%%%%%%%%%%%%%%%%%%%%%%%%%%%
%%%%%%%%%%%%%%%%%%%%%%%%%%%%%%%%%%%%%%%%%%
\begin{abstract}

We find stochastic equations governing eigenvalues and eigenvectors of a dynamical complex Ginibre ensemble reaffirming the intertwined role played between both sets of matrix degrees of freedom. We solve the accompanying Smoluchowski-Fokker-Planck equation valid for any initial matrix. We derive evolution equations for the averaged extended characteristic polynomial and for a class of $k$-point eigenvalue correlation functions. From the latter we obtain a novel formula for the eigenvector correlation function which we inspect for Ginibre and spiric initial conditions and obtain macro- and microscopic limiting laws.

\end{abstract}
%%%%%%%%%%%%%%%%%%%%%%%%%%%%%%%%%%%%%%%%%%

%02.10.Yn - Matrix theory
% 02.50.Ey - Stochastic Processes
%05.90.+m Other topics in statistical physics, thermodynamics, and nonlinear dynamical systems

%%%%%%%%%%%%%%%%%%%%%%%%%%%%%%%%%%%%%%%%%%%%%%%%%%%%%%%%%%%%%%%%%%%%%%

\maketitle

%%%%%%%%%%%%%%%%%%%%%%%%%%%%%%%%%%%%%%%%%%%%%%%%%%%%%%%%%%%%%%%%%%%%%%
%%%%%%%%%%%%%%%%%%%%%%%%%%%%%%%%%%%%%%%%%%%%%%%%%%%%%%%%%%%%%%%%%%%%%
\section{Introduction}

There are two key distinctions of non-normal matrices: their eigenvalues are complex and their eigenvectors are not orthogonal. When the matrix elements become random, we can consider statistical properties of both. Since the publication of the seminal work by Ginibre \cite{GG}, the eigenvalues of non-Hermitian, including non-normal random matrices have been widely studied. The eigenvectors however, only recently are becoming the focus of increased attention. The pioneering works in this area, \cite{CHME1} and \cite{CHME2}, concentrate on averages of the so called overlap matrix $O_{\alpha\beta}=\braket{L_\alpha |L_\beta}\braket{R_\beta |R_\alpha}$ (where  $\ket{L}$ and $\ket{R}$ are left and right eigenvectors respectfully), and in particular the eigenvector correlation function
\begin{align} \label{O1}
  O(z,\bar{z})= \frac{1}{N^2}\left< \sum_{\alpha}O_{\alpha\alpha}\delta(z-\lambda_\alpha)\right>, 
\end{align}
which turnes out to be related to a generalised version of the resolvent function in the large matrix size limit \cite{NOWAKNOER}.
We now know that this and related objects have some interesting and quite diverse applications. In particular, the instability of fluid flows \cite{TREFScience} and complex systems \cite{MAY}, as well as transient behaviour in the evolution of the latter \cite{Gtrancient} is related to the non-orthogonality of the eigenvectors of associated matrices. The overlap matrix plays a prominent role in the studies of scattering in chaotic systems \cite{SASO, CHSCATTER, FYSA, GKLMRS}, and formula (\ref{O1}) turns up as the so-called Peterman factor describing the linewidth of a laser cavity mode \cite{PETER1}. Finally, uncovering the properties of large, time-lagged correlation matrices also leads to tackling problems of correlations of non-orthogonal eigenvectors \cite{NT_timelag}.

Interestingly, shortly before Ginibre's paper, another field defining work was published by Dyson \cite{DYSONbm}. There, he proposed to consider a matrix with elements performing independent Brownian motions. Using perturbation theory, he derived a Smoluchowski-Fokker-Planck equation governing the evolution of the joint probability density function for the eigenvalues. This approach was more recently adapted to derive and solve partial differential equations describing the dynamics of the resolvent (or Greens function) and  the averaged characteristic polynomial for diffusing Hermitian \cite{BN1, BGNW0} and Wishart matrices \cite{BNW1, BNW2}. 
More importantly for this paper, this was extended to the study of complex non-Hermitian matrices \cite{BGNTW1, BGNTW2}, where the intimate interplay of the introduced correlation function \eqref{O1} and the spectral resolvent was uncovered in the form of a coupling of associated nonlinear partial differential equations. These were derived by inspecting dynamic properties of a novel, extended form of an associated averaged characteristic polynomial, which turned out to satisfy a very simple diffusion equation in an auxiliary spatial-like dimension.  

With the renewed interest, novel developments emerged and mixed-matrix moments of the Ginibre ensemble were related to the overlap function \cite{WS} and results for products of  independent complex Ginibre matrices were obtained \cite{BSV}. Additionally, the famous Haagerup-Larsen theorem was extended to contain an eigenvectors part \cite{NT_free1, NT_free2} and the diagrammatic method was employed to obtain new results on the eigenvector correlation functions \cite{NT_twopoint}.

In the mean time, matrix based stochastic processes of real, non-symmetric matrices were also studied. In this context the large matrix size limit of a two-time correlation function of spin variables associated with real eigenvalues was calculated in \cite{TZ}. Later, a similar stochastic evolution was shown to be governed by an effectively attractive force \cite{MOVAS}. 

Finally, in a very recent development, P. Bourgade and G. Dubach showed that the diagonal overlaps of the complex Ginibre ensemble are governed in the bulk by the inverse gamma distribution and provided many more, associated results \cite{BD}. Moreover, they provide the Langevin description of the Dyson-type dynamics for eigenvalues of non-normal matrices. They show, that when the complex matrix elements perform Brownian motion, the evolution of the eigenvalues is driven only through their dependence on the eigenvectors. However, they do not derive the equation associated with the eigenvector themselves. Simultaneously, Y. Fyodorov, using a different method, in particular the supersymmetry approach, derived the joint probability density functions for eigenvalues and the eigenvector correlation function for both real and complex, non-self-adjoint, Gaussian random matrices, analysing additionally its bulk and edge scaling limits \cite{FYOD}.

Here, we revisit the problem of complex diffusing non-Hermitian matrices. Our aim is to complete the associated picture by providing full Dyson evolution equations of both eigenvalues and eigenvectors. Thus, in section \ref{ch:Dys}, we start the paper by deriving these evolution equations. Section \ref{ch:SFPder} is devoted to the resulting Smoluchowski-Fokker-Planck partial differential equation as an alternative way for representing the dynamics. It contains both the derivation of the equation and its solution. In section \ref{ch:D}, we re-derive the above mentioned diffusion equation for the extended characteristic polynomial - this time using the result from section \ref{ch:Dys} of this paper. Finally, by exploiting the interplay of the evolution of eigenvalues and eigenvectors, in section \ref{ch:Corr}, we obtain results relating their correlation functions. These in turn provide a novel exact formula for the eigenvector correlation function \eqref{O1}. In many cases, the more complicated calculations are moved to the appendices. We finish the paper with some conclusions and acknowledgments. 

%%%%%%%%%%%%%%%%%%%%%%%%%%%%%%%%%%%%%%%%%%%%%%%%%%%%%%%%%%%%%%%%%%%%%%
%%%%%%%%%%%%%%%%%%%%%%%%%%%%%%%%%%%%%%%%%%%%%%%%%%%%%%%%%%%%%%%%%%%%%
\section{Stochastic evolution of non-Hermitian random matrices}
\label{ch:Dys}

Let us start our discussion by considering a general perturbation of a complex matrix $X$ of size $N\times N$. First, we introduce a similarity transformation:
\begin{align}
\label{transf}
X = S \Lambda S^{-1},
\end{align}
where $\Lambda$ is diagonal and contains the eigenvalues whereas the $S$ matrix encodes the eigenvectors with $S_{\alpha\beta} = \ket{R_\beta}_\alpha$. By It\^o calculus we find the differential:
\begin{align*}
\dd X = \dd S \Lambda S^{-1} + S \dd \Lambda S^{-1} + S \Lambda \dd (S^{-1}) + \dd S \dd \Lambda S^{-1} + S \dd \Lambda \dd (S^{-1}) + \dd S \Lambda \dd (S^{-1})
\end{align*}
expressed as
\begin{align}
\label{maineq}
\delta X = \delta S \Lambda + \dd  \Lambda + \Lambda \delta S' + \delta S \dd  \Lambda + \dd \Lambda \delta S' + \delta S \Lambda \delta S',
\end{align}
where $\delta X = S^{-1} \dd X S$, $\delta S = S^{-1} \dd S$ and $\delta S' = \dd (S^{-1}) S$. Notice that in general $\delta S' \neq (\delta S)^{-1}$. 

For readers not used to the formalism, we provide here a crude but sufficient crash-course to the calculus. We view every differential term $\dd\alpha = \dd_{\text{fv}}\alpha + \dd_{\text{m}} \alpha$ as containing two independent parts -- finite variance (denoted in this work by a subscript `$\text{fv}$') and martingale/stochastic (denoted by a subscript `$\text{m}$'). The latter is formally proportional to $\sqrt{\dd t}$ whereas the former is linear in the time increment $\dd t$. This mnemotechnical rule implies expanding Eq. \eqref{maineq} to the second order in differentials (unlike the usual differential calculus). Accordingly, to calculate the second order differentials $\dd\alpha \dd \alpha$ we treat both parts as usual numbers and use some straight-forward combining rules: $\dd_{\text{fv}} \dd_{\text{m}} \to 0, \dd_{\text{fv}} \dd_{\text{fv}} \to 0, \dd_{\text{m}} \dd_{\text{m}} \to \dd_{\text{fv}}$. 

Using these techniques we show in appendix \ref{a:ito} that the perturbation of $X$ implies infinitesimal changes of eigenvalues $\dd \Lambda$ and eigenvectors $\delta S$, encoded by the following equations: 
\begin{align}
 \dd \lambda_{i} & = \delta X_{ii} + \sum_{k(\neq i)} \frac{\delta X_{ik} \delta X_{ki}}{\lambda_i - \lambda_k}, \label{lambdas}\\
 (\delta S)_{ij} & = \frac{\delta X_{ij}}{\lambda_j - \lambda_i} + \sum_{k(\neq j)} \frac{\delta X_{ik} \delta X_{kj}}{(\lambda_i - \lambda_j)(\lambda_k - \lambda_j)} - \frac{\delta X_{ij} \delta X_{jj}}{(\lambda_j - \lambda_i)^2}, \qquad i\neq j, \label{deltas}\\
 (\delta S)_{ii} & = 0,
 \label{constraints}
\end{align}
where $(\dd \Lambda)_{ii} = \dd \lambda_i$ and the last equation is a set of $2N$ real constraints imposed on diagonal variations in order to retain the number of degrees of freedom $2N^2 = (2N^2-2N) + 2N$.

\subsection{Dyson's evolution}
%%%%%%%%%%%
Now, we will narrow our considerations to a simple stochastic process. Let the matrix $X_{ij}$ undergo a diffusion, with the real and imaginary parts of the elements evolving independently:
\begin{align}
\label{stochdef}
\dd X_{kl} = \sqrt{2} \left ( \dd B^{(1)}_{kl} + i \dd B^{(2)}_{kl} \right ),
\end{align}
where $\dd B^{(i)}_{kl}$ denotes a standard real Brownian motion. This differential contains only the martingale part $\dd X_{\text{m},kl}$ - the finite variance part $\dd X_{\text{fv},kl} $ vanishes. This form implies the second order variations $\dd X_{ij} \dd X_{kl} = 0, \quad \dd X_{ij} \dd \bar{X}_{kl} =  \delta_{ik} \delta_{jl} \dd t$. Since stochastic equations \eqref{lambdas}-\eqref{constraints} depend on $\delta X$, we find it useful to calculate also $\delta X_{ij} \delta \bar{X}_{kl} = (S^\dagger S)_{lj} (S^\dagger S)^{-1}_{ik} \dd t = A_{lj} A^{-1}_{ik} \dd t $ where we introduced $A=S^\dagger S$ related also to the overlap matrix $O_{ij} = A^{-1}_{ij} A_{ji}$ introduced in Eq. \eqref{O1}.

Such stochastic dynamics simplify considerabily the increments given by Eqs. \eqref{lambdas}-\eqref{constraints}. The Dyson's dynamics of eigenvalues and eigenvectors is given by:
\begin{align}
\dd \lambda_{ii} & = \delta X_{ii}, \label{lambdadelta}\\
\delta S_{ij} & = \frac{\delta X_{ij}}{\lambda_j - \lambda_i}, \qquad i \neq j, \label{sdelta}\\
\delta S_{ii} & = 0.\label{constrdelta}
\end{align}
One striking feature of the evolution is already evident -- the lack of a Vandermonde-like interaction term in the dynamics of eigenvalues. It is instead present in the dynamics of eigenvectors as first observed numerically in \cite{BlGNTW}. The relation of these equations with known non-Hermitian results will be established in section \ref{solsfp}, where a special solution to the Smoluchowski-Fokker-Planck equation associated with Dyson's evolution equation coincides with the complex Ginibre ensemble studied in \cite{GG}.

We now turn to including the constraint $\delta S_{ii} = 0$ into the evolution. We use $\delta S = S^{-1} \dd S$ and compute $\dd S_{ij}$:

\begin{align}
\sum_{i(\neq j)} \sum_k S_{li} S^{-1}_{ik} \dd S_{kj} = \sum_{i(\neq j)} \frac{S_{li} \delta X_{ij}}{\lambda_j - \lambda_i}.
\end{align}
From $S^{-1} S = 1$, one finds $\sum_{i(\neq j)} S_{li} S_{ik}^{-1} = \delta_{lk} - S_{lj} S^{-1}_{jk}$ which is plugged into the previous equation:
\begin{align}
\dd S_{lj} - S_{lj} \sum_k S^{-1}_{jk} \dd S_{kj} = \sum_{i(\neq j)} \frac{S_{li} \delta X_{ij}}{\lambda_j - \lambda_i}.
\end{align}
The second term on the l.h.s. vanishes $\sum_k S^{-1}_{jk} \dd S_{kj} = \delta S_{jj} = 0$ due to the constraint. The reduced Dyson's evolution is now described by
\begin{align}\label{Dysonmain1}
\dd \lambda_i & = \delta X_{ii}, \\ \label{Dysonmain2}
\dd S_{ij} & = \sum_{l(\neq j)} \frac{S_{il} \delta X_{lj}}{\lambda_j - \lambda_l},\quad \text{any} ~~i,j=1...N,
\end{align}
where now the constraint given by Eq. \eqref{constrdelta} is included implicitly in the dynamics. The (non-zero) second order differentials are found by the It\^o calculus methods:
\begin{align}
\dd \lambda_i \dd \bar{\lambda}_j & = A^{-1}_{ij} A_{ji} dt, \label{lambdalambda} \\
\dd S_{kl} \dd \bar{\lambda}_i & = \sum_{n(\neq l)} \frac{S_{kn}}{\lambda_l - \lambda_n} A_{il} A^{-1}_{ni} dt ,\label{dlambdadS}\\
\dd S_{kl} \dd \bar{S}_{nm} & = \sum_{\substack{\alpha(\neq l)\\ \beta(\neq m)}} \frac{S_{k\alpha} \bar{S}_{n\beta} A_{ml} A^{-1}_{\alpha\beta}}{(\lambda_l - \lambda_\alpha)(\bar{\lambda}_m - \bar{\lambda}_\beta)} dt \label{dSdS},
\end{align}
which completes the Dyson's evolution of both eigenvalues and eigenvectors. A distinct feature of the motion of eigenvalues alone is its variance $\dd \lambda_i \dd \bar{\lambda}_i = O_{ii} \dd t$ depending on the overlap matrix which necessarily speeds up the stochastic motion as $O_{ii}\geq 1$. A related phenomenon was observed in \cite{MOVAS} and Eq. \eqref{lambdalambda} was also derived recently in \cite{BD}. 

Finally, to highlight the significance of these equations, we juxtapose them with a two-dimensional Coulomb gas description. There, instead of solving the Dyson's problem, one starts from the steady Ginibre ensemble jPDF and infers a dynamical system producing the associated distribution. In that way, the following Langevin description of interacting eigenvalues arises naturally:
\begin{align}
\label{Coulomb}
\dd \lambda_j = \sqrt{2} \left ( \dd B^{(1)}_j +i \dd B^{(2)}_j \right ) - \frac{2}{N} \sum_{k(\neq j)} \frac{\lambda_k - \lambda_j}{|\lambda_k - \lambda_j|^2} \dd t -2\lambda_j \dd t ,
\end{align}
with $\braket{\dd B^{a}_j(t)}=0$ and $\braket{\dd B^{(a)}_{i}(t) \dd B^{(b)}_{j}(t')}=\frac{1}{2N}\delta(t-t')\delta_{ij}\delta_{ab}\dd t$.  This system of equations describes a 2D Coulomb-gas of eigenvalues \cite{BCFCOULOMB,OSADA} with a pairwise repulsive interaction and a harmonic potential added so that the stationary solution gathers the eigenvalues in a disk of radius one. Although the Dysonian approach given by Eqs. \eqref{Dysonmain1}-\eqref{Dysonmain2} and Coulomb-gas method of Eq. \eqref{Coulomb} are drastically different, by construction they give rise to the same ensemble for a special choice of vanishing initial conditions. In general, this difference stems from the fact that the Coulomb gas method can be reformulated as a Dyson-like approach for \textit{normal} \cite{CHAUZABOR,WZ2003:NORMALMATRIX} non-Hermitian random matrices i.e. a subset fulfilling the condition $[X,X^\dagger] = 0$. To shed more light on the striking difference between the evolution of the two-dimensional Coulomb gas and of non-Hermitian, non-normal matrices, we simulate the two and show the results in Fig 1. In the latter case the elements of the matrix (the $X_{ij}(t)$) undergo an Ornstein-Uhlenbeck process, such that $\braket{\dd X_{ij}(t)}=-\frac 14 X_{ij}(t)\dd t$ and $\braket{\dd X_{ij}(t) \dd\bar{ X}_{nm}(t')}=\frac{1}{2N}\delta(t-t')\delta_{in}\delta_{jm}\dd t$. The dynamics of the former is governed by Eq. (\ref{Coulomb}). The time steps are of exactly the same size in both simulations and the flow of time is captured by the increasing opacity of the disks representing the eigenvalues in the plots of the left panel. In both cases the global statistics are the same, as shown by the histograms in the right panels - a flat distribution of eigenvalues, confined to a disk in the complex plane, implies a semi-circular distribution of the real (or imaginary) parts of the eigenvalues. The microscopic dynamics is however very different. The 'particles' of the Coulomb gas behave similarly across the whole spectrum and their trajectories can be traced on the plot. The same can be said about most of the eigenvalues on the edge of the spectrum of the diffusing matrix, but as is well known, the closer to the center of the spectrum we get the larger the eigenvector correlator becomes. This results in an increase of the variance of the eigenvalue position (as according to Eq. (\ref{lambdalambda})), and thus, deep in the spectrum, the particular trajectories are not distinguishable.

\begin{figure}[htbp] \label{fig:stpr}
\centering
\includegraphics[width=0.98\textwidth]{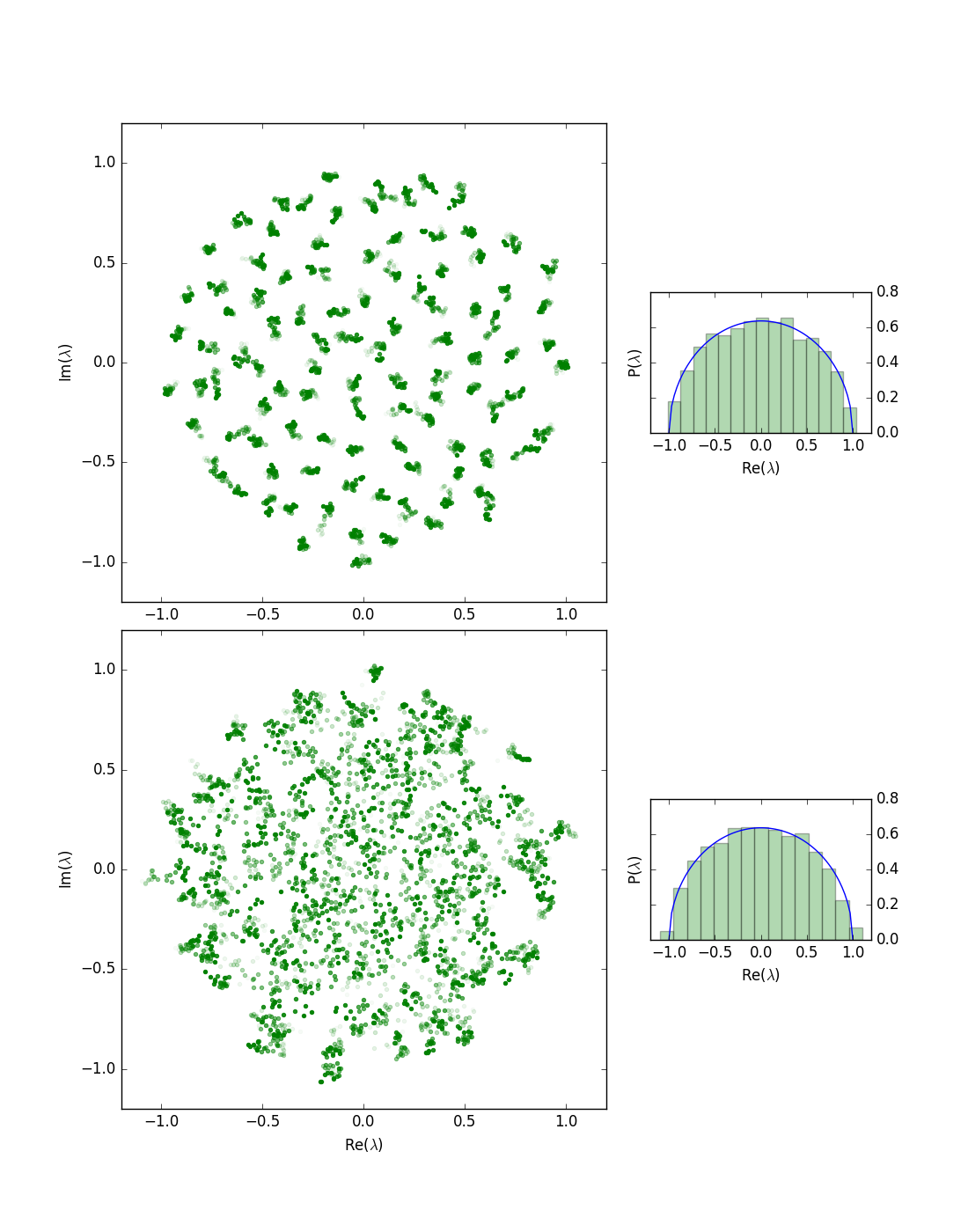}
\caption{Left panels: 30 time steps of the planar Coulomb gas evolution (upper panel) governed by Eq. \eqref{Coulomb} and an additive non-hermitian matrix evolution (lower panel) governed by Eq. \eqref{stochdef}, captured by increasing opacity of the disks marking the associated 100 eigenvalues. Right panels: histograms of the real parts of the eigenvalues for the former (upper panel) and latter (lower panel) evolution types (averaged over 40 time steps separated by 20 time step intervals). Both histograms converge to the same semi-circular law (stemming from the circular law) despite apparent differences in the dynamics of single-particle trajectories.}
\end{figure}

%%%%%%%%%%%%%%%%%%%%%%%%%%%%%%%%%%%%%%%%%%%%%%%%%%%%%%%%%%%%%%%%%%%%%%
%%%%%%%%%%%%%%%%%%%%%%%%%%%%%%%%%%%%%%%%%%%%%%%%%%%%%%%%%%%%%%%%%%%%%
\section{The Smoluchowski-Fokker-Planck equation}
\label{ch:SFPder}
Instead of looking at the Dyson's evolution, we can study the associated, deterministic Smoluchowski-Fokker-Planck (SFP) partial differential equation, describing the joint probability density function (jPDF) for the eigenvalues and eigenvectors.
To this end, we find it best to rewrite Eqs. \eqref{lambdadelta}-\eqref{constrdelta} as:
\begin{align*}
\dd \lambda_i & = \delta X_{ii}, \\
\dd  R_{ij} & = \frac{\delta X_{ij}}{\lambda_j - \lambda_i}, \quad i \neq j,\\
\dd R_{ii} & = 0,
\end{align*}
where we changed the notation $\delta S_{ij} \equiv \dd  R_{ij}$ in accordance with a differential character of this object $\delta S = S^{-1} \dd S$. We calculate the second order differentials and obtain:
\begin{align}
\dd \lambda_i \dd \bar{R}_{kl} & = \frac{A_{li} A^{-1}_{ik}}{\bar{\lambda}_l - \bar{\lambda}_k} \dd t, \label{dlambdadR}\\
\dd  R_{ij} \dd \bar{R}_{kl} & = \frac{A_{lj} A^{-1}_{ik}}{(\bar{\lambda}_l - \bar{\lambda}_k)(\lambda_j - \lambda_i)} \dd t. \label{dRdR}
\end{align}

Although there is no simple formula relating (non-infinitesimal) $R$ and $S$, we look for an equation for jPDF $P_t (R,\Lambda)$, depending on $R$, $\Lambda$ and time $t$. 
To this end, consider an arbitrary observable $f_t \equiv f(R(t),\Lambda(t))$ and compute its evolution equation 
\begin{align}
\label{ddf}
\dd f_t = \dd_{\text{fv}} f_t + \dd_{\text{m}} f_t,
\end{align}
where the finite variance and martingale differential operators read
\begin{align}
\dd_{\text{fv}} & = \sum_{i,j} \dd \lambda_i \dd \bar{\lambda}_j \partial_{\lambda_i \bar{\lambda}_j}  + \sum_{i,k\neq l} \dd \lambda_i \dd \bar{R}_{kl} \partial_{\lambda_i,\bar{R}_{kl}}  + \sum_{i,k\neq l} \dd \bar{\lambda}_i \dd R_{kl} \partial_{\bar{\lambda}_i,R_{kl}} + \sum_{k\neq l, n\neq m} \dd R_{kl} \dd \bar{R}_{nm} \partial_{R_{kl}, \bar{R}_{nm}}, \label{dfv}\\
\dd_{\text{m}} & = \sum_i d\lambda_i \partial_{\lambda_i} + \sum_i d\bar{\lambda}_i \partial_{\bar{\lambda}_i} + \sum_{k\neq l} dR_{kl} \partial_{R_{kl}} + \sum_{k\neq l} d\bar{R}_{kl} \partial_{\bar{R}_{kl}}.
\end{align}
This result follows from the It\^o's lemma, however it also coincides with a simple expansion of $f_t$ up to the order $\dd t$. The evolution equation \eqref{ddf} is mathematically equivalent to a sum of the usual Lebesgue (finite variance part) and stochastic (martingale part) integral $f_t = f_0 + \int_0^t \dd u \frac{ \dd_{\text{fv}}}{\dd u} f_u + \int_0^t \dd u \frac{ \dd_{\text{m}}}{\dd u} f_u$ where $f_0$ is the initial value of the observable. Next, we take its mean value:
\begin{align}
\label{E0}
\left < f_t \right > = f_0 + \int_0^t \dd u \left < \frac{ \dd_{\text{fv}}}{\dd u} f_u \right > + \int_0^t \dd u \left < \frac{ \dd_{\text{stoch}}}{\dd u} f_u \right >,
\end{align}
with averaging taken over the unknown jPDF as $\left < f_t \right > = \int [\dd R] [\dd \Lambda] P_t(R,\Lambda) f(R,\Lambda)$ and with entrywise measures $[\dd R] = \prod\limits_{i\neq j} \dd  R{ij} \dd \bar{R}_{ij}$, $[\dd \Lambda] = \prod\limits_i \dd \lambda_i \dd \bar{\lambda}_i$. The averaging makes the stochastic part vanish $ \left < \frac{ \dd_{\text{m}}}{\dd u} f_u \right >=0$. We now compute the partial derivative ($\partial_t$) of Eq. \eqref{E0} and write down the averages explicitly:

\begin{align*}
\int [\dd R] [\dd \Lambda] \partial_t P_t(R,\Lambda) f(R,\Lambda) = & \int [\dd R] [d\Lambda] f(R,\Lambda) \left [  \sum_{i,j} \partial_{\lambda_i \bar{\lambda}_j} \left ( \frac{\dd \lambda_i d\bar{\lambda}_j}{\dd t} P_t(R,\Lambda) \right ) + \sum_{i,k\neq l} \partial_{\lambda_i,\bar{R}_{kl}}\left ( \frac{d\lambda_i d\bar{R}_{kl}}{\dd t} P_t(R,\Lambda) \right ) + \right . \\
 & \qquad \qquad \qquad \qquad \left . + \sum_{i,k\neq l} \partial_{\bar{\lambda}_i,R_{kl}} \left ( \frac{\dd \bar{\lambda}_i \dd R_{kl}}{\dd t}  P_t(R,\Lambda) \right ) + \sum_{k\neq l, n\neq m} \partial_{R_{kl}, \bar{R}_{nm}} \left ( \frac{dR_{kl} \dd \bar{R}_{nm}}{\dd t} P_t(R,\Lambda) \right )  \right ],
\end{align*}
where we plugged in the finite variance differential operator given by Eq. \eqref{dfv} and integrated each term by parts, so that all derivatives act on the jPDF $P_t$. Because function $f$ is arbitrary, we skip the integrals and write down the SFP equation:
\begin{align}
\label{SFPR}
\partial_t P_t = & \sum_{i,j} \partial_{\lambda_i \bar{\lambda}_j} \left ( \frac{\dd \lambda_i \dd \bar{\lambda}_j}{\dd t} P_t \right ) + \sum_{i,k\neq l} \partial_{\lambda_i,\bar{R}_{kl}}\left ( \frac{\dd \lambda_i \dd \bar{R}_{kl}}{\dd t} P_t \right ) + \sum_{i,k\neq l} \partial_{\bar{\lambda}_i,R_{kl}} \left ( \frac{\dd \bar{\lambda}_i \dd R_{kl}}{\dd t}  P_t \right ) + \sum_{k\neq l, n\neq m} \partial_{R_{kl}, \bar{R}_{nm}} \left ( \frac{\dd R_{kl} \dd \bar{R}_{nm}}{dt}  P_t\right ).
\end{align}

The solution $P_t(R,\Lambda)$ is however given in terms of matrix $R$ whose exact relation to eigenvector-encoding $S$ is unknown. To aid this, we change the variables $R \to S$ in the SFP equation \eqref{SFPR}. The transformation is tractable since the following identities hold
\begin{align}
\sum_{k\neq l} \frac{\partial}{\partial \bar{R}_{kl}} \left ( \frac{\dd \lambda_i \dd \bar{R}_{kl}}{\dd t} P_t \right ) & = \sum_{k, l} \frac{\partial}{\partial \bar{S}_{kl}} \left ( \frac{\dd \lambda_i \dd \bar{S}_{kl}}{\dd t} P_t \right ), \label{id1}\\
\sum_{k\neq l,n\neq m} \frac{\partial}{\partial R_{kl} \partial \bar{R}_{nm}} \left ( \frac{\dd R_{kl} d\bar{R}_{nm}}{\dd t} P_t \right ) & = \sum_{k, l,n,m} \frac{\partial}{\partial S_{kl} \partial \bar{S}_{nm}} \left ( \frac{\dd S_{kl} \dd \bar{S}_{nm}}{\dd t} P_t \right ). \label{id2}
\end{align}
To show this we find the transformed differential operator
\begin{align}
\label{dr}
\frac{\partial}{\partial R_{ij}} = \sum_{nm} \frac{\dd S_{nm}}{\dd R_{ij}} \frac{\partial}{\partial S_{nm}} = \sum_n S_{ni} \frac{\partial}{\partial S_{nj}}
\end{align}
by the definition $S\dd R = \dd S$ which entails $\sum_k S_{nk} \dd R_{km} = \dd S_{nm}$ and $ \frac{\dd S_{nm}}{\dd R_{ij}} = S_{ni} \delta_{mj}$. We plug Eqs. \eqref{dr} and \eqref{dlambdadR} on the l.h.s. of identity \eqref{id1}:
\begin{align*}
\sum_{k\neq l} \frac{\partial}{\partial \bar{R}_{kl}} \left ( \frac{\dd \lambda_i \dd \bar{R}_{kl}}{\dd t} P_t \right ) = \sum_{k\neq l} \sum_n \bar{S}_{nk} \frac{\partial}{\partial \bar{S}_{nl}} \left ( \frac{A_{li}A^{-1}_{ik}}{\bar{\lambda}_l - \bar{\lambda}_k} P_t \right ) = \sum_{l,n} \frac{\partial}{\partial \bar{S}_{nl}}  \left ( \sum_{k (\neq l)}\frac{\bar{S}_{nk} A_{li}A^{-1}_{ik}}{\bar{\lambda}_l - \bar{\lambda}_k} P_t \right ) = \sum_{l,n} \frac{\partial}{\partial \bar{S}_{nl}} \left ( \frac{\dd \lambda_i \dd \bar{S}_{nl}}{dt} P_t \right ),
\end{align*}
which recreates the r.h.s. of Eq. \eqref{id1}. We have used differential operator $\bar{S}_{nk} \partial_{\bar{S}_{nl}} = \partial_{\bar{S}_{nl}} \bar{S}_{nk}  - \delta_{nn} \delta_{kl}$ and the fact that we sum over $k\neq l$. Analogous computation can be done also for the second identity \eqref{id2}.

Using these identities, the $S$-dependent SFP equation reads:
\begin{align}
\label{SFPS}
\partial_t P_t = & \sum_{i,j} \partial_{\lambda_i \bar{\lambda}_j}\left(\frac{ \dd \lambda_i \dd \bar{\lambda}_j}{\dd t} P_t \right ) + \sum_{i,k,l} \partial_{\lambda_i,\bar{S}_{kl}}\left ( \frac{\dd \lambda_i \dd \bar{S}_{kl}}{\dd t} P_t \right ) + \sum_{i,k, l} \partial_{\bar{\lambda}_i,S_{kl}} \left ( \frac{\dd \bar{\lambda}_i \dd S_{kl}}{\dd t}  P_t \right ) + \sum_{k,l, n,m} \partial_{S_{kl}, \bar{S}_{nm}} \left ( \frac{\dd S_{kl} \dd \bar{S}_{nm}}{\dd t}  P_t \right ),
\end{align}
where $P_t \equiv P_t(S,\Lambda)$.

%%%%%%%%%%%%%%%%%%%%%%%%%%%%%%%%%%%%%%%%%%%%%%%%%%%
\subsubsection{Solutions to SFP equation \eqref{SFPS}}
\label{solsfp}

We turn to investigate concrete solutions to the SFP equation \eqref{SFPS}. We first introduce a succinct notation of each term:
\begin{align}
\label{SFPS2}
\partial_t P_t = & \sum_{i,j} C_{ij}^{\lambda,\bar{\lambda}} \partial_{\lambda_i \bar{\lambda}_j} P_t + \sum_{i,k,l} \partial_{\lambda_i,\bar{S}_{kl}}\left ( C_{ikl}^{\lambda,\bar{S}} P_t \right ) + \sum_{i,k, l} \partial_{\bar{\lambda}_i,S_{kl}} \left ( C_{ikl}^{S,\bar{\lambda}} P_t \right ) + \sum_{k,l, n,m} \partial_{S_{kl}, \bar{S}_{nm}} \left ( C_{klnm}^{S,\bar{S}} P_t \right ),
\end{align}
where $C_{ij}^{\lambda,\bar{\lambda}} = A^{-1}_{ij} A_{ji} $, $C_{ikl}^{\lambda,\bar{S}} = \sum\limits_{n(\neq l)} \frac{\bar{S}_{kn} A_{li} A^{-1}_{in}}{\bar{\lambda}_l - \bar{\lambda}_n} $, $C_{ikl}^{S,\bar{\lambda}} = \sum\limits_{n(\neq l)} \frac{S_{kn} A_{il} A^{-1}_{ni}}{\lambda_l - \lambda_n}$ and $C_{klnm}^{S,\bar{S}} = \sum\limits_{\substack{\alpha (\neq l),\\ \beta (\neq m)}} \frac{S_{k\alpha} \bar{S}_{n\beta} A_{ml} A^{-1}_{\alpha\beta}}{(\lambda_l - \lambda_\alpha)(\bar{\lambda}_m -\bar{\lambda}_\beta)} $.
 We look for a solution in the following separated form:
\begin{align}
\label{ansatz}
P_t(S,\Lambda) = F(\Lambda) Q_t (S,\Lambda).
\end{align}
We plug this ansatz into Eq. \eqref{SFPS2} and split the terms on the r.h.s. into three groups:
\begin{align}
\text{r.h.s. of \eqref{SFPS2}} = T_Q + T_{\partial Q} + T_{\partial\partial Q},
\end{align}
where $T_Q$ gathers all terms proportional to $Q_t$, $T_{\partial Q}$ th terms depending on the first derivative of $Q_t$ and $T_{\partial\partial Q}$ all the terms containing second derivatives. We write down $T_Q$:
\begin{align}
\label{TQ}
T_Q & = Q_t \left [ \sum_{i,j} C_{ij}^{\lambda,\bar{\lambda}} \partial_{\lambda_i \bar{\lambda}_j} F + \sum_{i,k,l}\partial_{\bar{S}_{kl}}C_{ikl}^{\lambda,\bar{S}} \partial_{\lambda_i}F + \sum_{i,k,l}\partial_{S_{kl}}C_{ikl}^{S,\bar{\lambda}} \partial_{\bar{\lambda}_i}F + \sum_{k,l, n,m} \partial_{S_{kl}, \bar{S}_{nm}} C_{klnm}^{S,\bar{S}} F \right ].
\end{align}
In appendix \ref{ap:identities2} we show that for a particular choice of the proportionality factor:
\begin{align}
F(\Lambda) = \prod_{i<j} |\lambda_j - \lambda_i|^4,
\end{align}
the terms in the bracket vanish resulting in $T_Q = 0$. Similarly, when we look at $T_{\partial Q}$:
\begin{align}
\label{TdQ}
T_{\partial Q} & =  \sum_{j} \partial_{\bar{\lambda}_j} Q_t \left [ \sum_i C_{ij}^{\lambda,\bar{\lambda}} \partial_{\lambda_i } F + F \sum_{k,l}\partial_{S_{kl}}C_{jkl}^{S,\bar{\lambda}} \right ]  + \sum_{i}  \partial_{\lambda_i } Q_t \left [ \sum_j C_{ij}^{\lambda,\bar{\lambda}} \partial_{\bar{\lambda}_j} F + F\sum_{k,l}\partial_{\bar{S}_{kl}}C_{ikl}^{\lambda,\bar{S}} \right ] + \\
& + \sum_{k,l} \partial_{S_{kl}} Q_t \left [ \sum_i C_{ikl}^{S,\bar{\lambda}} \partial_{\bar{\lambda}_i} F +  F \sum_{n,m} \partial_{\bar{S}_{nm}} C_{klnm}^{S,\bar{S}} \right ] + \sum_{k,l} \partial_{\bar{S}_{kl}} Q_t \left [ \sum_i C_{ikl}^{\lambda,\bar{S}} \partial_{\lambda_i} F + F \sum_{nm} \partial_{S_{nm}} C_{nmkl}^{S,\bar{S}}\right ],
\end{align}
we find in appendix \ref{ap:identities2} that it also vanishes $T_{\partial Q}=0$. Only the last term survives
\begin{align*}
T_{\partial \partial Q} & = F \left [ \sum_{i,j} C_{ij}^{\lambda,\bar{\lambda}} \partial_{\lambda_i \bar{\lambda}_j} Q_t + \sum_{i,k,l} C_{ikl}^{\lambda,\bar{S}} \partial_{\lambda_i,\bar{S}_{kl}} Q_t + \sum_{i,k,l} C_{ikl}^{S,\bar{\lambda}} \partial_{\bar{\lambda}_i,S_{kl}} Q_t +  \sum_{k,l, n,m} C_{klnm}^{S,\bar{S}} \partial_{S_{kl}, \bar{S}_{nm}} Q_t \right ],
\end{align*}
and combined with $\partial_t P_t = F \partial Q_t$ gives the SFP equation for $Q_t$:
\begin{align}
\label{SFPQ}
\partial_t Q_t = \sum_{i,j} C_{ij}^{\lambda,\bar{\lambda}} \partial_{\lambda_i \bar{\lambda}_j} Q_t + \sum_{i,k,l} C_{ikl}^{\lambda,\bar{S}} \partial_{\lambda_i,\bar{S}_{kl}} Q_t + \sum_{i,k,l} C_{ikl}^{S,\bar{\lambda}} \partial_{\bar{\lambda}_i,S_{kl}} Q_t +  \sum_{k,l, n,m} C_{klnm}^{S,\bar{S}} \partial_{S_{kl}, \bar{S}_{nm}} Q_t.
\end{align}
We find a solution of the form:
\begin{align}
\label{Qtsol}
Q_t = \frac{c}{t^{N^2}} \exp \left ( -\frac{1}{t} \sum_{i,j,k,l} \left ( S^{\dagger,-1}_{ik} \bar{\lambda}_k S^\dagger_{kj} - S^{(0),\dagger,-1}_{ik} \bar{\lambda}_k^{(0)} S^{(0),\dagger}_{kj}\right ) \left ( S_{jl} \lambda_l S^{-1}_{li} - S^{(0)}_{jl} \lambda_l^{(0)} S^{(0),-1}_{li} \right ) \right ),
\end{align}
where the fixed eigenvalues and eigenvectors are respectively denoted by $\lambda^{(0)}$ and $S^{(0)}$ and there is some time-independent constant $c$. It is an initial value solution as in the $t\to 0$ limit we find
\begin{align*}
\lim_{t \to 0} Q_t \sim \prod_{i,j} \delta^{(2)} \left (X_{ij} - X^{(0)}_{ij}\right),
\end{align*}
with $X^{(0} = S^{(0} \Lambda^{(0)} S^{(0),-1}$. We read-off the full jPDF from Eq. \eqref{ansatz} as:
\begin{align*}
P_t(S,\Lambda) [\dd \Lambda] [\dd R] = \frac{1}{(2\pi t)^{N^2}N!} \prod_{i<j} |\lambda_j-\lambda_i|^4 \exp \left ( -\frac{1}{t} \sum_{i,j,k,l} \left ( S^{\dagger,-1}_{ik} \bar{\lambda}_k S^\dagger_{kj} - S^{(0),\dagger,-1}_{ik} \bar{\lambda}_k^{(0)} S^{(0),\dagger}_{kj}\right ) \left ( S_{jl} \lambda_l S^{-1}_{li} - S^{(0)}_{jl} \lambda_l^{(0)} S^{(0),-1}_{li} \right ) \right ) [\dd \Lambda] [\dd R], 
\end{align*}
where $[\dd \Lambda] = \prod_i \dd \lambda_i \dd \bar{\lambda}_i, [\dd R] = \prod_{i\neq j} \dd R_{ij}$, $\dd R = S^{-1} \dd S$. This general form can be also deduced based on the derivations of chapter $15$ in \cite{Meh2004:RMTBOOK}.
The constant $c = \left ((2\pi)^{N^2} N! \right )^{-1}$ is found by considering the special case $X^{(0)} = 0$, where the solution $Q_t$ simplifies:
\begin{align}
Q_t\left (X^{(0)} = 0 \right ) = \frac{1}{(t)^{N^2}} \exp \left ( -\frac{1}{t} \sum_{i,j} A^{-1}_{ij} A_{ji} \lambda_i \bar{\lambda}_j \right ),
\end{align}
and the jPDF:
\begin{align}
P_t\left (S,\Lambda; X^{(0)}=0 \right ) [\dd \Lambda] [\dd R] = \frac{1}{(2\pi t)^{N^2}N!} \prod_{i<j} |\lambda_j-\lambda_i|^4 \exp \left (-\frac{1}{t} \sum_{i,j} A^{-1}_{ij} A_{ji} \lambda_i \bar{\lambda}_j \right ) [\dd \Lambda] [\dd R]
\end{align}
is a form found by Ginibre in his seminal paper \cite{GG}. The fourth power of the Vandermonde term found in the context of complex matrices can be surprising at first but one should remember that typically the second power of Vandermonde appears when one considers a marginal jPDF with eigenvector variables integrated out. Indeed, this task was completed by Ginibre himself and the eigenvector integration yields the Vandermonde to power $-2$ resulting in the well-known result.

%%%%%%%%%%%%%%%%%%%%%%%%%%%%%%%%%%%%%%%%%%%%%%%%%%%%%%%%%%%%%%%%%%%%%%
%%%%%%%%%%%%%%%%%%%%%%%%%%%%%%%%%%%%%%%%%%%%%%%%%%%%%%%%%%%%%%%%%%%%%
\section{The characteristic polynomial}
\label{ch:D}
As we have mentioned in the introduction, it was shown in earlier works, that the so-called Extended Characteristic Polynomial (ECP)
\begin{align}
\label{Ddef}
D \equiv \det \left [ (z-X)(\bar{z} - X^\dagger) + |w|^2 \right ] = \det \left [ \begin{matrix} z - \Lambda & - \bar{w} A^{-1} \\ w A & \bar{z} - \Lambda^\dagger \end{matrix} \right ] \equiv \det M
\end{align}
gives access to the bulk properties of non-Hermitian random matrices in the large $N$ limit \cite{BGNTW1}. In particular, after averaging over the elements of $X$ (denoted by $\left< \right >$) it can be used to calculate the spectral density, with 
\[
\rho(z,\bar{z})=\lim_{N \to \infty}\lim_{|w|\to 0} \frac{1}{\pi N}\del_{z\bar{z}}\left <\log D\right >
\]
and the one point eigenvector overlap function defined by Eq. \eqref{O1}, with
\[
O(z,\bar{z})=\lim_{N \to \infty}\lim_{|w|\to 0} \frac{1}{\pi N^2}\left| \del_{w}\left <\log  D\right > \right|^2,
\]
in some sense analogically to the Hermitian case \cite{BlGNTW} and where, to our knowledge the self-averaging property in the large $N$ limit $\log\left <  D\right >=\left < \log D\right >$ is not formally proven yet. For the stochastic process under consideration (defined by Eq. \eqref{stochdef}), it fulfils the following, remarkably simple and exact partial differential equation
\begin{align}\label{ddyn}
\partial_t \left < D \right > = \partial_{w\bar{w}} \left < D \right>,
\end{align} 
true for any size of the matrix and initial condition (note that time variable $t = \tau/N$ is rescaled in \cite{BGNTW1}). It can be solved both for finite $N$ and in the limit of the infinite size of the matrix. Originally, it was derived with the use of the diffusion equation satisfied by the matrix elements and by representing the determinant as a Grassmann variable integral. To show the usefulness of the full Dyson's equations governing the evolution of eigenvalues and eigenvectors introduced in the previous section, we will now employ them to re-derive Eq. (\ref{ddyn}).

First, let us introduce some notation. We represent the matrix in Eq. \eqref{Ddef} as $M = \left ( \begin{matrix}
a & b \\ c & d
\end{matrix} \right ),
$ where $a = z-\Lambda$, $b = -\bar{w} A^{-1}$, $c = w A$ and $d = \bar{z} - \Lambda^\dagger$. The inverse is in turn given by $M^{-1} = \left ( \begin{matrix}
M'_{11} & M'_{12} \\
M'_{21} & M'_{22}
\end{matrix}
  \right )$ where $M'_{11} = \xi, M'_{12} = - \xi b d^{-1}, M'_{21} = - d^{-1} c \xi$, $M'_{22} = (d- c a^{-1} b)^{-1} = b^{-1} a \xi b d^{-1}$ and $\xi = (a - b d^{-1} c)^{-1}$. 
We use It\^o formula and calculate:
\begin{align}\label{dD}
\dd D = \dd_{\text{m}} D + \sum_{i,j} \dd \lambda_i \dd \bar{\lambda}_j \partial_{\lambda_i,\bar{\lambda}_j} D + \sum_{i,k,l} \dd \lambda_i \dd \bar{S}_{kl} \partial_{\lambda_i,\bar{S}_{kl}} D + \sum_{k,l,j} \dd S_{kl} \dd \bar{\lambda}_j \partial_{S_{kl},\bar{\lambda}_j} D + \sum_{k,l,n,m} \dd S_{kl} \dd \bar{S}_{nm} \partial_{S_{kl},\bar{S}_{nm}} D.
\end{align}
Since we aim at computing the diffusion equation for $\left < D \right >$, the stochastic or martingale part is not crucial as it will be averaged out (see also Sec. \ref{ch:SFPder}). 
A straight-forward but lenghty calculation gives
\begin{align*}
\dd D = \dd_{\text{m}}D + D {\left[ \Tr (b^{-1} a \xi b d^{-1}A \xi A^{-1}) - \Tr (d^{-1} c\xi A^{-1})\Tr (\xi b d^{-1} A) \right]} ~\dd t.
\end{align*}
It turns out, that $D^{-1} \partial_{w\bar{w}} D $ gives precisely the inside of the $[\cdots]$ brackets in the second term of the right-hand side of this equation. Thus, after averaging out the stochastic term, we prove Eq. (\ref{ddyn}).

%%%%%%%%%%%%%%%%%%%%%%%%%%%%%%%%%%%%%%%%%%%%%%%%%%%%%%%%%%%%%%%%%%%%%%
%%%%%%%%%%%%%%%%%%%%%%%%%%%%%%%%%%%%%%%%%%%%%%%%%%%%%%%%%%%%%%%%%%%%%%
\section{Relations between eigenvalue and eigenvector correlation functions}
\label{ch:Corr}

Let us now use the evolution equations \eqref{ddf} to derive relations between eigenvalue and eigenvector correlation functions. We focus on the $k$-point eigenvalue correlation functions of the form:
\begin{align*}
\hat{\rho}_k(z_1,...,z_k) & = \sum_{i_1\neq i_2, \neq,\cdots, i_k} \prod_{n=1}^k \delta (z_n - \lambda_{i_n}),
\end{align*}
among which, the two first examples are the (not normalized) spectral density $\hat{\rho}_1(z) = \sum_{i=1}^N \delta(z-\lambda_i)$ and the (not normalized) two-point correlation function $\hat{\rho}_2(z_1,z_2) = \sum_{i\neq j}^N \delta(z_1-\lambda_i)\delta(z_2-\lambda_j)$.
We derive simple dynamical equations for these quantities as they do not depend on eigenvectors:
\begin{align}
\label{stocheq}
\dd \hat{\rho}_k = \sum_{\alpha,\beta=1}^k \partial_{z_\alpha\bar{z}_\beta} \hat{O}_k(z_1, \hat{z}_\alpha,...,\hat{z}_\beta,...,z_k) \dd t + \dd_{\text{m}} \hat{\rho}_k,
\end{align}
where $\dd_{\text{m}}$ denotes the non-essential stochastic part and $\hat{O}_k$ is a generalized $k$-eigenvector correlation function:
\begin{align*}
\hat{O}_k(z_1, \hat{z}_\alpha,...,\hat{z}_\beta,...,z_k) \equiv \hat{O}_k(\alpha,\beta) = \sum_{i_1\neq i_2, \neq,\cdots, i_k} O_{i_\alpha i_\beta} \prod_{n=1}^k \delta (z_n - \lambda_{i_n}).
\end{align*}
In the $k=1$ case, the definition is proportional to the eigenvector correlation function given by Eq. \eqref{O1} as $\hat{O}_1(\hat{\hat{z}}_1) = \sum_i O_{ii} \delta (z_1 - \lambda_i)$. The averaged version of stochastic equation \eqref{stocheq} is given in terms of normalized objects $\rho_k = \frac{1}{N^k} \left <\hat{\rho}_k \right >$ and $O_k(\alpha,\beta) = \frac{1}{N^k} \left < \hat{O}_k(\alpha,\beta) \right >$ and with a rescaled time variable $t = \tau/N$:
\begin{align}
\label{hierarchy}
\partial_\tau \rho_k = \frac{1}{N} \sum_{\alpha,\beta=1}^k \partial_{z_\alpha\bar{z}_\beta} O_k(\alpha,\beta) ,
\end{align}
which remains valid for any matrix size $N$. We continue to study this family of equations first in the case $k=1$ where both $O_1,\rho_1$ are familiar and leave the general $k>1$ case for future research.

%%%%%%%%%%%%%%%%%%%%%
\subsection{The case of $k=1$}
We set $k=1$, denote $\rho_1 \equiv \rho$ as the normalized spectral density and $O = \frac{1}{N} O_1$ as the eigenvector correlation function normalized as in the definition \eqref{O1}. A particularly simple relation between eigenvector and eigenvalue related quantities reads
\begin{align}
\label{0hierarchy}
\partial_\tau \rho = \partial_{z\bar{z}} O.
\end{align}
First, we simply cross-check it against the known bulk formulas for the initial matrix $X_0 = 0$ where spectral density and eigenvector correlator are $\rho \sim \frac{1}{\pi \tau}$, $O \sim \frac{\tau - |z|^2}{\pi \tau^2}$ respectively and remain valid inside a circle of radius $|z|=\sqrt{\tau}$. In the large $N$ limit, the same relation \eqref{0hierarchy} holds also for a general class of bi-unitarily invariant non-Hermitian random matrices as was shown recently in \cite{NT_free1}. 

Since Eq. \eqref{0hierarchy} is exact, we focus on obtaining novel finite $N$ results from this relation. At first glance, we cannot simply solve one equation with two unknown functions however once we have one of them, the second can be deduced rather easily. In \cite{HP1998:NHEXTDENSITY}, the spectral density formula valid for finite matrix size $N$ and a normal initial matrix $X_0$ was found:
\begin{align}
\label{pniniform}
\rho(z,\bar{z}) = \frac{1}{\pi N} \int_0^1 \dd\beta ~ \partial_{z\bar{z}} \left [ \oint_{\mathcal{C}} \frac{\dd v}{2\pi i} \frac{e^{-\beta \frac{N}{\tau} v}}{\beta^2 v} \prod_{i=1}^N \left ( 1 - \frac{\beta v}{v- |a_i-z|^2} \right )  \right ],% \quad \text{[CH3]}
\end{align}
where $X_0 = \text{diag} (a_1...a_N)$ and the contour $\mathcal{C}$ encircles counter-clockwise poles located at $|a_i - z|^2$. Although this expression was obtained in the framework of external source non-Hermitian models, there exists a simple connection to the diffusive approach discussed in this paper. Namely, the external source corresponds to the matrix initiating the evolution, whereas the overall variance of the random matrix is proportional to the evolution time $\tau$. 
Most importantly, the second derivative $\partial_{z\bar{z}}$ seems to fit in the equation \eqref{0hierarchy} neatly. First however, we must address a subtlety hidden in the expression \eqref{pniniform} -- the integral over $\beta$ and the derivative operator $\partial_{z\bar{z}}$ cannot commute as taking the $\beta$ integral first leads to a divergent answer due to the $1/\beta^2$ factor (or $1/\beta$ if we calculate the contour integral $v$ first). We signal this problem now and will propose a solution shortly. Application of equation \eqref{0hierarchy} to obtain a formula for $O_1$ is therefore at first stated formally in terms of an inverse derivative operator $(\partial_{z\bar{z}})^{-1}$:
\begin{align*}
O(z,\bar{z}) = \frac{1}{\pi N} (\partial_{z\bar{z}})^{-1} \partial_\tau \int_0^1 \dd \beta ~ \partial_{z\bar{z}} \left [ \oint_{\mathcal{C}} \frac{\dd v}{2\pi i} \frac{e^{-\beta \frac{N}{\tau} v}}{\beta^2 v} \prod_{i=1}^N \left ( 1 - \frac{\beta v}{v- |a_i-z|^2} \right )  \right ].
\end{align*}
First, we find $\beta$ integral and $\tau$ derivative commuting %[CH4]
which can be understood since rendering the $\beta$-integral convergent demands taking at least one derivative:
\begin{align*}
O(z,\bar{z}) = \frac{1}{\pi \tau^2} (\partial_{z\bar{z}})^{-1} \int_0^1 \dd \beta ~ \partial_{z\bar{z}} \left [ \oint_{\mathcal{C}} \frac{\dd v}{2\pi i} \frac{e^{-\beta \frac{N}{\tau} v}}{\beta} \prod_{i=1}^N \left ( 1 - \frac{\beta v}{v- |a_i-z|^2} \right )  \right ],
\end{align*}
which cures the divergence as we demonstrate now. We compute the $v$ integral:
\begin{align*}
\left [ \cdots \right ] = \oint_{\mathcal{C}} \frac{\dd v}{2\pi i} \frac{e^{-\beta \frac{N}{\tau} v}}{\beta} \prod_{i=1}^N \left ( 1 - \frac{\beta v}{v- |a_i-z|^2} \right ) = \oint_{\mathcal{C}} \frac{\dd v}{2\pi i} \frac{e^{-\beta \frac{N}{\tau} v}}{\beta} \left ( 1 + \beta f(v,\beta) \right ) = \oint_{\mathcal{C}} \frac{\dd v}{2\pi i} e^{-\beta \frac{N}{\tau} v} f(v,\beta)
\end{align*}
where $f(v,\beta) = -\sum_i \frac{v}{v-|a_i-z|^2} + \sum_{i\neq j} \frac{ \beta v^2}{(v-|a_i-z|^2)(v - |a_j-z|^2)} + \cdots + \frac{(-1)^N\beta^{N-1} v^N}{\prod_{i=1}^N (v- |a_i-z|^2)}$. The culprit term proportional to $1/\beta$ vanishes upon calculating the contour integral first, before taking the $\beta$ integral. Because $f$ is a finite polynomial in $\beta$ variables, its $\beta$ integral is also finite. We can now safely commute the derivatives %[CH4]
and obtain the formula:
\begin{align*}
O(z,\bar{z}) = \frac{1}{\pi \tau^2} (\partial_{z\bar{z}})^{-1} \partial_{z\bar{z}} \int_0^1 \dd \beta ~  \left [ \oint_{\mathcal{C}} \frac{\dd v}{2\pi i} e^{-\beta \frac{N}{\tau} v} f(v,\beta) \right ] = \frac{1}{\pi \tau^2} (\partial_{z\bar{z}})^{-1} \partial_{z\bar{z}} \int_0^1 \dd \beta ~  \left [ \oint_{\mathcal{C}} \frac{dv}{2\pi i} \frac{e^{-\beta \frac{N}{\tau} v}}{\beta} \prod_{i=1}^N \left ( 1 - \frac{\beta v}{v- |a_i-z|^2} \right ) \right ] .
\end{align*}
Lastly, we set $(\partial_{z\bar{z}})^{-1} \partial_{z\bar{z}} = 1$ and add a function $C$ satisfying $\partial_{z\bar{z}} C=0$:
\begin{align*}
O(z,\bar{z}) = \frac{1}{\pi \tau^2} \int_0^1 \dd \beta ~ \left [ \oint_{\mathcal{C}} \frac{\dd v}{2\pi i} \frac{e^{-\beta \frac{N}{\tau} v}}{\beta} \prod_{i=1}^N \left ( 1 - \frac{\beta v}{v- |a_i-z|^2} \right ) \right ] + C,
\end{align*}
which is found with a natural assumption of $O(z,\bar{z}) \to 0$ as $z \to \infty$ and assuming that function $C$ is bounded and therefore a constant from the Liouville's theorem. Since we are interested in large $z$ asymptotics, without loosing any generality, we set $a_i = 0$ and compute the integral
\begin{align}
\label{I}
I(z) = \int_0^1 \dd \beta \oint_{|z|^2} \frac{\dd v}{2\pi i} \frac{e^{-\beta \frac{N}{\tau} v}}{\beta} \left ( 1 - \frac{\beta v}{v- |z|^2} \right )^N = \sum_{k=1}^N \binom{N}{k} \frac{(-1)^k}{(k-1)!} \frac{\dd ^{k-1}}{\dd v^{k-1}} \left [ v^k \int_0^1 \dd \beta \beta^{k-1} e^{-\beta \frac{N}{\tau} v} \right ]_{v=|z|^2} ,% \quad \text{[CH5]}
\end{align}
where we have used the binomial formula and the residue theorem. Using the integral $\int_0^1 d\beta e^{-\beta a} \beta^k = \frac{k!-\Gamma(k+1,a)}{a^{k+1}}$ %[CH6]
splits $I(z) = I_{\text{const}} + I_{\text{var}}(z)$ %[CH7]
into two parts:
\begin{align}
I_{\text{const}} & = \sum_{k=1}^N \binom{N}{k} \left ( - \frac{\tau}{N} \right )^k \frac{\dd^{k-1}}{\dd v^{k-1}} \left [ 1 \right ]_{v=|z|^2} = -\tau, \\ %\quad \text{[CH8]}
I_{\text{var}}(z) & = -  \sum_{k=1}^N \binom{N}{k} \frac{1}{(k-1)!} \left ( -\frac{\tau}{N} \right )^k \frac{\dd^{k-1}}{\dd v^{k-1}} \left [ \Gamma\left (k, \frac{Nv}{\tau} \right ) \right ]_{v=|z|^2}. \label{I1}
\end{align}
We immediately obtain $I_{\text{var}}(z) \to 0$ as $z\to \infty$ %[CH9]
due to the behaviour of the incomplete Gamma function. Since the eigenvector correlator $O = \frac{I}{\pi \tau^2} + C$ ought to vanish in this limit and so from $0 = \frac{I_{\text{const}}}{\pi \tau^2} + C$ we arrive at $C = \frac{1}{\pi \tau}$. The final exact formula for the eigenvector correlator reads:
\begin{align}
\label{O1fin}
O(z,\bar{z}) = \frac{1}{\pi \tau^2} \int_0^1 \dd \beta ~ \left [ \oint_{\mathcal{C}} \frac{\dd v}{2\pi i} \frac{e^{-\beta \frac{N}{\tau} v}}{\beta} \prod_{i=1}^N \left ( 1 - \frac{\beta v}{v- |a_i-z|^2} \right ) \right ] + \frac{1}{\pi \tau}. %\quad \text{[CH10]}
\end{align}
The formula is valid for normal initial matrix $X_0 = \text{diag}(a_1...a_N)$. For non-normal matrix $X_0$, the relation \eqref{0hierarchy} is still valid however exact results for the spectral density are limited (see \cite{GG2016:EXACTSPECTR} for rank-one non-normal deformation). 
%%%%%%%%%%%%%%%%%%%%%
\subsubsection{Ginibre case} We compute the expression \eqref{O1fin} in the classic case of $a_i=0$, recreating the Ginibre ensemble. The eigenvector correlation function reads
\begin{align*}
O_{\text{Ginibre}}(z,\bar{z}) = \frac{1}{\pi \tau^2} \int_0^1 \dd \beta ~ \left [ \oint_{\mathcal{C}} \frac{\dd v}{2\pi i} \frac{e^{-\beta \frac{N}{\tau} v}}{\beta} \left ( 1 - \frac{\beta v}{v- |z|^2} \right )^N \right ] + \frac{1}{\pi \tau} = \frac{1}{\pi \tau^2} I(z) + \frac{1}{\pi \tau} = \frac{1}{\pi \tau^2} I_{\text{var}}(z),
\end{align*}
with integrals $I(z)$ and $I_{\text{var}}(z)$ computed previously in Eqs. \eqref{I} and \eqref{I1} respectively. We transform the latter term further:
\begin{align}
\label{Ivar}
I_{\text{var}}(z) = \frac{\tau}{N} \sum_{m=0}^{N-1} \frac{\Gamma(m+1,N|z|^2/\tau)}{m!} = \frac{\tau}{N} e^{- \frac{N|z|^2}{\tau}} \sum_{m=0}^{N-1} \frac{N-m}{m!} \left (\frac{N|z|^2}{\tau} \right )^m, %\quad \text{[CH11]}
\end{align}
The resulting eigenvector correlator:
\begin{align}
\label{eqqq}
O_{\text{Ginibre}}(z,\bar{z}) = \frac{1}{\tau \pi N} e^{- \frac{N|z|^2}{\tau}} \sum_{m=0}^{N-1} \frac{N-m}{m!} \left (\frac{N|z|^2}{\tau} \right )^m,
\end{align}
agrees with the well-known result \cite{WS}. For completeness, we show below both macro- and microscopic limiting laws arising from Eq. \eqref{eqqq}.

\paragraph{Macroscopic bulk regime.} We calculate first how the macroscopic limit is restored. To this end, let $z \sim O(1)$ and we approximate the sum of \eqref{eqqq} with the Euler-Maclaurin formula. We set $ m = N \alpha$ and use the Stirling expansion $1/(N\alpha)! \sim \frac{e^{N\alpha (1-\log \alpha - \log N)}}{\sqrt{2\pi N \alpha}}$ to find the leading term in terms of an integral $\sum_{m=0}^{N-1} \frac{N-m}{m!} \left (\frac{N|z|^2}{\tau} \right )^m \sim \frac{N^2}{\sqrt{2\pi N}} \int_0^1 \dd \alpha \frac{1-\alpha}{\sqrt{\alpha}} e^{N f(\alpha)}$ where $f(\alpha) = \alpha \left ( 1 + \log \left ( |z|^2/\tau \right ) - \log \alpha \right )$. The integral is found by standard saddle-point method as $\int_0^1 \dd \alpha \frac{1-\alpha}{\sqrt{\alpha}} e^{N f(\alpha)} \sim \frac{\sqrt{2\pi}e^{\frac{N|z|^2}{\tau}}}{\tau \sqrt{N}} \left ( \tau - |z|^2 \right )$ when $|z|^2<\tau$ and $\sim \frac{e^N (\frac{|z|^2}{\tau})^N}{N^2 (\log |z|^2/\tau)^2}$ otherwise.

Gathering the above formulas gives the classic macroscopic result \cite{CM1998:EIGENVECTOR}:
\begin{align}
\lim_{N \to \infty} O_{\text{Ginibre}}(z,\bar{z}) = \frac{1}{\pi \tau^2} (\tau - |z|^2) \theta (\tau - |z|^2). % \quad \text{[CH19]}
\end{align}

\paragraph{Microscopic edge regime.} In the microscopic limit we zoom near the edge of the density by setting $|z| = \sqrt{\tau} + \delta N^{-1/2}$. Instead of using formula \eqref{eqqq}, we derive an alternative integral representation of $I_{\text{var}}$ given by Eq. \eqref{Ivar}, namely:
\begin{align}
\label{Ivarintegr}
I_{\text{var}} = (-1)^{N+1} \int_0^\infty \dd u e^{-\frac{N}{\tau} u} (u+|z|^2)^N \frac{1}{2\pi i} \oint_{C(|z|^2)} \dd \sigma \frac{e^{-\frac{N}{\tau}\sigma}}{u+\sigma} \frac{1}{(\sigma - |z|^2)^N}, %\qquad \text{[CH39]}
\end{align}
The $\sigma$ integral is approximated by saddle-point method around $\sigma = |z|^2-\tau - i s N^{-1/2}$:
\begin{align*}
& (-1)^N \frac{1}{2\pi i} \oint_{C(|z|^2)} \dd \sigma \frac{e^{-\frac{N}{\tau}\sigma}}{u+\sigma} \frac{1}{(\sigma - |z|^2)^N} \sim - \frac{1}{2} \tau^{-N} e^{-2\delta/\tau \sqrt{N}} e^{\frac{(\sqrt{N} u + 2 \delta \sqrt{\tau})^2}{2\tau^2}} \text{erfc} \left ( \frac{\sqrt{N} u + 2 \delta \sqrt{\tau}}{\tau \sqrt{2}} \right ), %\quad \text{[CH37]}
\end{align*}
The remaining $u$ integral in the expression \eqref{Ivarintegr} is computed similarly, by expanding $u = (- 2\delta \sqrt{\tau}+ v) N^{-1/2}$, resulting in $I_{\text{var}} \sim \frac{1}{2\sqrt{N}} \int_{2\delta \sqrt{\tau}}^\infty \dd v ~\text{erfc} \left ( \frac{v}{\sqrt{2} \tau} \right )$.
With the definition $O_{\text{Ginibre}} = \frac{1}{\pi \tau^2} I_{\text{var}}$, we obtain the edge microscopic law found also in \cite{WS}:
\begin{align}
\label{edgemicro}
\lim_{N \to \infty} \sqrt{N} O_{\text{Ginibre}}\left (|z| = \sqrt{\tau} + \delta N^{-1/2} \right ) = \frac{1}{\pi \tau} \left ( \frac{1}{\sqrt{2\pi}} e^{\frac{-2\delta^2}{\tau}} - \frac{\delta}{\sqrt{\tau}}  \text{erfc} \left (\frac{\sqrt{2} \delta}{\sqrt{\tau}} \right )  \right ). %\quad \text{[CH22]}
\end{align}
The eigenvector correlation function thus behaves smoothly near the edge following an error-function-type universality law akin to those found in the edge microscopic regime for the spectral density.

%%%%%%%%%%%%%%%%%%%%%
\subsubsection{Spiric case} 
\label{spiric}
We consider now a prime example of non-vanishing initial matrix $X_0$. Let $N$ be an even number and we set the initial positions to $a_i= (-1)^i a$. Thus, initially the eigenvalues are positioned (in equal amounts) at symmetric points $\pm a$. This special choice is known in the literature as the spiric case and was studied in the bulk limit (or large $N$) in \cite{BGNTW2,FZ1997:HMETHOD} as the simplest instance in which there occurs a dynamical change in the topology of the spectral boundary. As the two initial bunches of eigenvalues spread, they are bound to collide at some finite time and create one eigenvalue bulk. To show that, we present the macroscopic correlation function $O$ is given by:
\begin{align}
\label{spiricmacro}
\lim_{N\to \infty} O_{\text{spiric}} = \begin{cases} \frac{1}{2\pi \tau^2} \left ( \tau - A_- - A_+ + \sqrt{\tau^2 + (A_+ - A_-)^2} \right ) & z \in C \\ 0 & z \notin C \end{cases}, 
\end{align}
where $A_\pm = |a\pm z|^2$ and $C$ is a contour defined by the spiric equation $\frac{\tau}{2} \left ( A_+ + A_- \right ) = A_+ A_-$. Spiric sections undergo a topological change at time $\tau_c = |a|^2$ when two initial eigenvalue islands centered around $\pm a$ start to coalesce at the origin $z=0$. The space-time point $(z,\tau) = (0,\tau_c)$ is therefore a pertinent candidate in search for new universal behaviour. 

The expression \eqref{O1fin} for the correlation function is again split as $O_{\text{spiric}} = \frac{1}{\pi \tau} + \frac{1}{\pi \tau^2} I_{\text{spiric}}$, where the integral reads
\begin{align}
\label{Ispiric}
I_{\text{spiric}} & = \int_0^1 \dd \beta ~ \left [ \oint_{\mathcal{C}} \frac{\dd v}{2\pi i} \frac{e^{-\beta \frac{N}{\tau} v}}{\beta} \left ( 1 - \frac{\beta v}{v- A_-} \right )^{N/2} \left ( 1 - \frac{\beta v}{v- A_+} \right )^{N/2} \right ].
 \end{align} 
Unfortunately, in present form, this expression is not very handy in conducting saddle-point computations, however we present now an alternative formulation which does not have such shortcomings: 
\begin{align}
\label{Ispiric2}
I_{\text{spiric}} & = - \tau - \int_0^\infty  \frac{\dd u}{2\pi i} \oint_{C(A_-,A_+)} d\sigma \frac{e^{-\frac{N}{\tau} (u +\sigma)}}{u+\sigma} \left ( \frac{(A_+ + u)(A_- + u)}{(\sigma-A_-)(\sigma-A_+)} \right )^{N/2}.
\end{align}
This version of the formula was derived in appendix \ref{reprspiric}. We will now study its behaviour in both macro- and microscopic regimes.

\paragraph{Macroscopic bulk regime.} We first calculate the macroscopic limit where $z \sim O(1)$ and so $A_\pm \sim O(1)$. The asymptotic form of the $\sigma$ contour integral in Eq. \eqref{Ispiric2} is computed first:
\begin{align*}
J_\sigma = \frac{1}{2\pi i} \oint_{C(A_-,A_+)} \dd \sigma \frac{e^{-\frac{N}{\tau} \sigma}}{u+\sigma} \left ( \frac{1}{(\sigma-A_-)(\sigma-A_+)} \right )^{N/2} =  J_\sigma^+ + J_\sigma^-,
\end{align*}
where we split the contour integral into two parts encircling only one of the poles. Depending on the sign of $A_+ - A_-$, one of the contributions will dominate. The geometric meaning of this difference is straightforward as $A_\pm$ measure the squared distance from the point $z$ to the source term located at $\pm a$. The sign of $A_+ - A_-$ therefore depends on which point is nearest to our probe located at $z$. We assume from now on that $A_- < A_+$ (or that we are in the vicinity of $a$) and so $J_\sigma^- \sim O(1)$ is leading wrt. $J_\sigma^+ \sim O(N^{-1/2})$ (it would be the opposite if $A_- > A_+$). The saddle points for both integrals are the same and given by $\sigma_*^\pm = \frac{1}{2} (A_- + A_+ - \tau \pm \sqrt{\tau^2 + (A_+ - A_-)^2})$. Crucially, they do not depend on the exchange $A_- \leftrightarrow A_+$ and therefore the saddle $\sigma_*^-$ will always be closer to the origin $\sigma=0$ than $\sigma_*^+$. That is, although the poles move around in the $\sigma$ space as we probe different areas of the complex plane $z$, the saddles themselves maintain a hierarchy and so always $\sigma_*^- < \sigma_*^+$. This means that if $A_-<A_+$ then contours centered around $A_\pm$ pick up the saddles at $\sigma_*^\pm$ but in the opposite case $A_->A_+$ the poles $A_\pm$ switch their relevant saddles and so pick up instead $\sigma_*^\mp$. Importantly, always the $\sigma_*^-$ saddle brings in the leading contribution as it is the only saddle capable of reaching negative half-space $\text{Re} \sigma<0$ and make contact with the pole at $-u$. We present this mechanism by writing down symbolically all the contributions depending on whether the saddle point $\sigma_*^-$ crosses or not the origin $\sigma=0$ and whether it interferes with the pole at $\sigma = -u$:
\begin{align*}
J_\sigma^- \sim \theta(\sigma_*^-)\oint_{C(A_-)}  + \theta(-\sigma_*^-) \left [ \theta(\sigma_*^- + u) \oint_{C(A_-)} + \theta(-\sigma_*^- - u) \left ( \frac{1}{2\pi i} \oint_{C(A_-, -u)}  - \frac{1}{2\pi i} \oint_{C(-u)}  \right )\right ].
\end{align*}
All saddle-point contributions besides the contour integral around $-u$ inevitably produce terms $O(N^{-1/2})$. The leading order is thus associated with a simple pole at $\sigma=-u$ and so $J_\sigma^- \sim - \theta(-\sigma_*^-) \theta(-\sigma_*^- - u) e^{\frac{N}{\tau} u} \left ( \frac{1}{(u+A_-)(u+A_+)} \right )^{N/2} $.
We plug it back into expression \eqref{Ispiric2} resulting in $I_{\text{spiric}} = - \tau + \theta(-\sigma_*^-) \int_0^{-\sigma_*^-} \dd u = - \tau - \sigma_*^-$ which, by relation $O_{\text{spiric}} = \frac{1}{\pi \tau} + \frac{1}{\pi \tau^2} I_{\text{spiric}}$, recreates the macroscopic formula \eqref{spiricmacro}:
\begin{align}
\label{Ospiric}
O_{\text{spiric}} = \frac{1}{\pi \tau^2} (-\sigma_*^-) \theta(-\sigma_*^-),
\end{align}
where $\sigma_*^- = \frac{1}{2} \left (A_- + A_+ - \tau - \sqrt{\tau^2 + (A_+ - A_-)^2} \right )$.

\paragraph{Microscopic collision regime.} 
We lastly turn to a microscopic regime near the collision time $\tau_c = |a|^2$ where both boundaries of spectral density and eigenvector correlator change its topology. We probe $I_{\text{spiric}}$ given by Eq. \eqref{Ispiric2} around the critical point in both space $z =\eta N^{-1/4}$ and time $\tau = |a|^2 + t N^{-1/2}$.
We start from approximating the $u$ integral:
\begin{align*}
\int_0^\infty \dd u \frac{e^{-\frac{N}{\tau} u}}{u+\sigma} \left [ (a_+ + u)(a_- + u)\right ]^{N/2} \sim e^{ g} \int_{-T_*}^\infty \frac{\dd v}{\sqrt{N}\sigma + T_* + v} e^{-\frac{v^2}{2|a|^4}},
\end{align*}
where $T_* = t+\frac{(a\bar{\eta} + \bar{a} \eta)^2-|a|^2|\eta|^2}{|a|^2}$ and $g = N\log |a|^2 -\frac{\sqrt{N}}{2}(\eta^2/a^2 + \bar{\eta}^2/\bar{a}^2) + \frac{t^2}{2|a|^4} + t\frac{(a\bar{\eta} + \bar{a} \eta)^2-|a|^2|\eta|^2}{|a|^6} + \frac{(\bar{a} \eta + a \bar{\eta})^4}{4|a|^8}$. Since the integral cannot be computed explicitly, we take only the $\sigma$ dependent part and calculate the $\sigma$ contour integral as
\begin{align*}
\frac{1}{2\pi i} \oint_{C(A_-,A_+)} \dd \sigma  \frac{e^{-\frac{N}{\tau} \sigma}}{\sqrt{N}\sigma + T_* + v} \frac{1}{\left ( (\sigma-A_-)(\sigma-A_+) \right )^{N/2}} \sim -\frac{e^{-g} }{2 \sqrt{N}} e^{\frac{v^2}{2|a|^4}} \text{erfc} \left ( \frac{v}{\sqrt{2}|a|^2}\right ). % \quad \text{[CH42]}
\end{align*}
Both formulas are combined and pre-factors cancel exactly resulting in $I_{\text{spiric}} \sim -\tau + \frac{1}{2\sqrt{N}} \int_{-T_*}^\infty \dd v ~\text{erfc}\left ( \frac{v}{\sqrt{2}|a|^2}\right )$.
Finally, we recall the relation $O_{\text{spiric}} = \frac{1}{\pi \tau} + \frac{1}{\pi \tau^2} I_{\text{spiric}}$ and obtain the collision microscopic law:
\begin{align}
\lim_{N\to \infty} \sqrt{N} O_{\text{spiric}}(z = \eta N^{-\frac{1}{4}}; \tau = |a|^2 + t N^{-1/2}) = \frac{1}{\pi |a|^2}\left ( \frac{1}{\sqrt{2\pi}} e^{-\frac{T_*^2}{2|a|^4}} + \frac{T_*}{2|a|^2} \text{erfc}\left ( - \frac{T_*}{\sqrt{2}|a|^2} \right ) \right ),
\end{align}
where $T_* = t+\frac{(a\bar{\eta} + \bar{a} \eta)^2-|a|^2|\eta|^2}{|a|^2}$. In the limit $\eta,\bar{\eta} \to \infty$, the formula agrees with the corresponding bulk equation \eqref{spiricmacro}. With identification $T_* \to -2\delta \sqrt{\tau}, |a|^2 \to \tau$, it has the same functional form as the Ginibre edge microscopic law \eqref{edgemicro}. Therefore, only the argument function $T_*$ encodes the spiric collision geometry but, besides that, the error-function-type universality class is not altered by a special choice of initial condition. 

\subsubsection{General case}
Based on previously studied examples, we propose a general formula for the eigenvector correlation function (alternative to Eq. \eqref{O1fin} derived earlier):
\begin{align}
\label{Ogeneral}
O = - \frac{1}{\pi \tau^2} \int_0^\infty \dd u \frac{1}{2\pi i} \oint_{C(\{A_i\})} \dd \sigma \frac{e^{-\frac{N}{\tau} (\sigma + u)}}{\sigma+u} \prod_{i=1}^N\frac{A_i + u}{A_i-\sigma}, %\qquad \text{[CH40]}
\end{align}
where $A_i = |a_i - z|^2$ and the contour integral encircles all $A_i$'s. In the special cases $a_i = 0$ and $a_i = (-a)^i$ it is reduced to formulas \eqref{Ivarintegr} and \eqref{Ispiric2}. Although the formula was not proved beyond two cases studied before, we have checked symbolically its validity against the representation given by Eq. \eqref{O1fin}. Its correctness is corroborated by the fact that the procedure of obtaining the macroscopic limiting law (as described in section \ref{spiric} for the spiric case) has a natural generalization to arbitrary sources. The result is again proportional to the smallest saddle point $\sigma_*^{\min}$ (compare with Eq. \eqref{Ospiric}):
\begin{align}
O = \frac{1}{\pi \tau^2} \theta(-\sigma_*^{\min})(-\sigma_*^{\min}),
\end{align}
where the minimal saddle is found from the equation $0 = -\frac{1}{\tau} + \frac{1}{N} \sum_{i=1}^N \frac{1}{A_i - \sigma}$. This simple result was already obtained in \cite{BGNTW2} but its interpretation was not clear. The equivalence can be shown by a simple notation change -- rename $\sigma = -r^2$, recall that $A_i = |a_i - z|^2$ and rewrite both the eigenvector correlator and equation for $\sigma_*^{\min} = -r_*^2$. Then we find
\begin{align*}
O = \frac{1}{\pi \tau^2} r_*^2, \qquad \frac{1}{\tau} = \frac{1}{N} \sum_{i=1}^N \frac{1}{r_*^2 + |a_i - z|^2},
\end{align*}
which are exactly Eqs. (39) and (40) of \cite{BGNTW2} and the $\theta(r_*^2)$ is implicitly present in these equations. In particular, the boundaries of the correlation functions are found by the condition $r_*=0$.

Lastly, we point out that the formula \eqref{Ogeneral} has a structure similar to the corresponding expressions in the Hermitian random matrices as found by Br\'ezin and Hikami \cite{BH1996:BHPAPER1} where two integral representations are largely independent and joined together only by a term $\frac{1}{u+\sigma}$. This suggests possible structures of the same kind arising also for higher--order correlation functions.

%%%%%%%%%%%%%%%%%%%%%%%%%%%%%%%%%%%%%%%%%%%%%%%%%%%%%%%%%%%%%%%%%%%%%%
%%%%%%%%%%%%%%%%%%%%%%%%%%%%%%%%%%%%%%%%%%%%%%%%%%%%%%%%%%%%%%%%%%%%%%
\section{Conclusions}
\label{ch:Conc}

In this work, we have tackled the problem of complex diffusing non-Hermitian matrices and provided a corresponding description of the behavior of their eigenvalues and eigenvectors via both stochastic differential equations and the Smoluchowski-Fokker-Planck equation. This allowed us to recover recent results, in particular the equation describing the diffusion of an extended characteristic polynomial and obtain a novel one, namely the form of an exact one point eigenvector correlation function for Ginibre matrices with an external source. We found both macro- and microscopic limiting laws for the case of vanishing and spiric-type source and showed that the error-function-type universality class encapsulates also a collision-type scenario, as was already hinted in the previous work \cite{BGNTW2}.

As the pace of the study field of non-Hermitian and in particular non-normal random matrices increases, we expect the appearance of more and more of new results and applications. We hope this work to be part of the prelude to that. For example, the subtle connections between the stochastic processes of eigenvalues and free entropy (and free information) for Hermitian matrices (see \cite{BLOWER}), make us hope that the presented advancements can be a stepping stone towards understanding the later notions for non-normal random matrices. Proposed integral representations for the eigenvector correlation function ought to be extended to spectral density where results are scarce. On the application side, we look toward stability problems, network science and artificial neural networks, however we expect to once more be surprised.
%%%%%%%%%%%%%%%%%%%%%%%%%%%%%%%%%%%%%%%%%%%%%%%%%%%%%%%%%%%%%%%%%%%%%%
%%%%%%%%%%%%%%%%%%%%%%%%%%%%%%%%%%%%%%%%%%%%%%%%%%%%%%%%%%%%%%%%%%%%%%
\section{Acknowledgments}
\label{ch:Ackn}

We would like to thank Roger Tribe and Oleg Zaboronski with whom we collaborated on the subject of the paper in its early stages. We are grateful for the hospitality of the Mathematics Institute of the Warwick University during that time. Furthermore, we would like to appreciate Maciej A. Nowak's and Wojciech Tarnowski's comments and an ongoing collaboration on topic related  to this work. PW expresses his gratitude to the Laboratorie de Physique Theorique et Modelles Statistiques (LPTMS CNRS, Paris-Sud) for hosting him when the last part of this project was realized. Finally, PW would like to acknowledge the funding of the Polish National Science Centre through the projects SONATA number 2016/21/D/ST2/01142.

%%%%%%%%%%%%%%%%%%%%%%%%%%%%%%%%%%%%%%%%%%%%%%%%%%%%%%%%%%%%%%%%%%%%%%
%%%%%%%%%%%%%%%%%%%%%%%%%%%%%%%%%%%%%%%%%%%%%%%%%%%%%%%%%%%%%%%%%%%%%%

%%%%%%%%%%%%%%%%%%%%%%%%%%%%%%%%%%%%%%%%%%%%%%%%%%%
%%%%%%%%%%%%%%%%%%%%%%%%%%%%%%%%%%%%%%%%%%%%%%%%%%%
%%%%%%%%%%%%%%%%%%%%%%%%%%%%%%%%%%%%%%%%%%%%%%%%%%%
\appendix
%%%%%%%%%%%%%%%%%%%%%%%%%%%%%%%%%%%%%%%%%%%%%%%%%%%
%%%%%%%%%%%%%%%%%%%%%%%%%%%%%%%%%%%%%%%%%%%%%%%%%%%
%%%%%%%%%%%%%%%%%%%%%%%%%%%%%%%%%%%%%%%%%%%%%%%%%%%

\section{Ito calculus} \label{a:ito}

We are interested in how a small perturbation of $X$ results in a changes $\dd \lambda$ and $\delta S$. We start from repeating Eq. \eqref{maineq}:
\begin{align*}
\delta X = \delta S \Lambda + \dd  \Lambda + \Lambda \delta S' + \delta S \dd  \Lambda + \dd \Lambda \delta S' + \delta S \Lambda \delta S'
\end{align*}
and read out its martingale part
\begin{align}
\label{deltaX}
\delta X_{\text{m}} = \delta S_{\text{m}} \Lambda + \dd\Lambda_{\text{m}} - \Lambda \delta S_{\text{m}}.
\end{align}
where $\delta S_{\text{m}} = - \delta S'_{\text{m}}$ was used from an identity:
\begin{align}
\label{ssp}
\delta S_ + \delta S' + \delta S \delta S' = 0,
\end{align}
found by considering a differential of $S^{-1}S = 1$. The diagonal part of Eq. \eqref{deltaX} is given by
\begin{align}
\label{dlm}
\dd \lambda_{{\text{m}},ii} = \delta X_{\text{m},ii}.
\end{align}
The off-diagonal part is equal to
\begin{align*}
\delta X_{ij} = (\lambda_j - \lambda_i) \delta S_{{\text{m}},ij},
\end{align*}
where we have used $\dd \Lambda_{\text{m},ij} = 0$. This results in
\begin{align}
\label{dsm}
\delta S_{{\text{m}},ij} = \frac{\delta X_{ij}}{\lambda_j - \lambda_i}, \qquad i\neq j.
\end{align}
The diagonal part $\delta S_{\text{m},ii}$ is arbitrary and we choose to set it to zero
\begin{align}
\label{constraint}
\delta S_{ii} = 0.
\end{align}
The finite variation part of \eqref{maineq} is:
\begin{align}
\label{deltaSfv}
\delta X_{\text{fv}} = \delta S_{\text{fv}} \Lambda + \dd\Lambda_{\text{fv}} + \Lambda \delta S'_{\text{fv}} + (\delta S \dd \Lambda)_{\text{fv}} + (\dd\Lambda \delta S')_{\text{fv}} + (\delta S \Lambda \delta S')_{\text{fv}}.
\end{align}
We use the rules of combining It\^o differentials to compute the following partial identities:
\begin{align*}
(\delta S \dd\Lambda)_{\text{fv},ij} & = \sum_{k\neq i} \delta S_{\text{m},ik} \dd\lambda_{\text{m},kj} = \frac{\delta X_{ij} \delta X_{jj}}{\lambda_j - \lambda_i} , \qquad i \neq j, \\
(\delta S \dd \Lambda)_{\text{fv},ii} & = \sum_{k} \delta S_{\text{m},ik} \dd\lambda_{\text{m},ki} = \delta S_{\text{m},ii} \delta X_{ii} = 0, \\
(\dd \Lambda \delta S')_{\text{fv},ij} & = \sum_{k\neq j} \dd \lambda_{\text{m},ik} \delta S'_{\text{m},kj} = \frac{\delta X_{ii} \delta X_{ij}}{\lambda_i - \lambda_j}, \qquad i \neq j,\\
(\dd\Lambda \delta S')_{\text{fv},ii} & = \sum_{k} \dd\lambda_{\text{m},ik} \delta S'_{\text{m},ki} = \delta S'_{\text{m},ii} \delta X_{ii} = 0, \\
(\delta S \Lambda \delta S')_{\text{fv},ij} & = \sum_{x\neq i,j} \delta S_{\text{m},ix} \lambda_x \delta S'_{M,xj} = \sum_{x \neq i,j} \frac{\lambda_x \delta X_{ix} \delta X_{xj}}{(\lambda_x - \lambda_i)(\lambda_x - \lambda_j)}, \\
(\delta S \delta S')_{\text{fv},ij} & = \sum_{k\neq i,j} \delta S'_{\text{m},ik} \delta S_{\text{m},kj} = \sum_{k\neq i,j} \frac{\delta X_{ik} \delta X_{kj}}{(\lambda_i - \lambda_k)(\lambda_j - \lambda_k)}, 
\end{align*}
From Eq. \eqref{ssp} we compute the finite variance part
\begin{align}
\label{ssp0}
\delta S_{\text{fv}} + \delta S'_{\text{fv}} + (\delta S \delta S')_{\text{fv}} = 0
\end{align}
and find 
\begin{align*}
\delta S'_{\text{fv},ij} & = - \delta S_{\text{fv},ij} - \sum_{k\neq i,j} \frac{\delta X_{ik} \delta X_{kj}}{(\lambda_i - \lambda_k)(\lambda_j - \lambda_k)}. 
\end{align*}

We move to computing the diagonal finite variation part of Eq. \eqref{deltaSfv} as
\begin{align*}
0 = \underline{\delta S_{\text{fv},ii}} \lambda_i + \dd \lambda_{\text{fv},ii} + \lambda_i \delta S'_{\text{fv},ii} + \underline{(\delta S \dd\Lambda)_{\text{fv},ii}} + \underline{(\dd \Lambda \delta S')_{\text{fv},ii}} + (\delta S \Lambda \delta S')_{\text{fv},ii},
\end{align*}
since $\delta X_{\text{fv}} = 0$ and the underlined parts vanish. We find that
\begin{align}
\dd \lambda_{\text{fv},ii} & = - \lambda_i \delta S'_{\text{fv},ii} - (\delta S \Lambda \delta S')_{\text{fv},ii} = \lambda_i (\delta S \delta S')_{\text{fv},ii} - (\delta S \Lambda \delta S')_{\text{fv},ii} = \sum_{k\neq i} \frac{\delta X_{ik} \delta X_{ki}}{\lambda_i - \lambda_k}. \label{dlfv}
\end{align}
In turn, the off-diagonal part reads:
\begin{align*}
0 = \delta S_{\text{fv},ij} \lambda_j + \underline{\dd\lambda_{\text{fv},ij}} + \lambda_i \delta S'_{\text{fv},ij} + (\delta S \dd \Lambda)_{\text{fv},ij} + (\dd \Lambda \delta S')_{\text{fv},ij} + (\delta S \Lambda \delta S')_{\text{fv},ij}, \qquad i \neq j,
\end{align*}
where the underlined term vanishes. With help of Eq. \eqref{ssp0}, we obtain:
\begin{align*}
\delta S_{\text{fv},ij} (\lambda_i - \lambda_j) & = - \lambda_i (\delta S \delta S')_{\text{fv},ij} + (\delta S \dd \Lambda)_{\text{fv},ij} + (\dd\Lambda \delta S')_{\text{fv},ij} + (\delta S \Lambda \delta S')_{\text{fv},ij} = \sum_{k\neq j} \frac{\delta X_{ik} \delta X_{kj}}{\lambda_k - \lambda_j} + \frac{\delta X_{ij} \delta X_{jj}}{\lambda_j - \lambda_i}.
\end{align*}
Thus, finally
\begin{align}
\label{dsfv}
(\delta S)_{\text{fv},ij} = \sum_{k\neq j} \frac{\delta X_{ik} \delta X_{kj}}{(\lambda_i - \lambda_j)(\lambda_k - \lambda_j)} - \frac{\delta X_{ij} \delta X_{jj}}{(\lambda_j - \lambda_i)^2}, \quad i \neq j.
\end{align}
We collect the differentials $\dd \lambda_{ii} = \dd \lambda_{\text{m},ii} + \dd \lambda_{\text{fv},ii}$ given by Eqs. \eqref{dlm} and \eqref{dlfv} and $\delta S_{ij} = \delta S_{\text{m},ij} + \delta S_{\text{fv},ij}$ given by Eqs. \eqref{dsm} and \eqref{dsfv}:
\begin{align*}
\dd \lambda_{ii} & = \delta X_{ii} + \sum_{k\neq i} \frac{\delta X_{ik} \delta X_{ki}}{\lambda_i - \lambda_k}, \\
\delta S_{ij} & = \frac{\delta X_{ij}}{\lambda_j - \lambda_i} + \sum_{k\neq j} \frac{\delta X_{ik} \delta X_{kj}}{(\lambda_i - \lambda_j)(\lambda_k - \lambda_j)} - \frac{\delta X_{ij} \delta X_{jj}}{(\lambda_j - \lambda_i)^2}, \qquad i \neq j, \\
\delta S_{ii} & = 0,
\end{align*}

which recreates Eqs. \eqref{lambdas}-\eqref{constraints}.

%%%%%%%%%%%%%%%%%%%%%%%%%%%%%%%%%%%%%%%%%%%%%%%%%%%
%%%%%%%%%%%%%%%%%%%%%%%%%%%%%%%%%%%%%%%%%%%%%%%%%%%
%%%%%%%%%%%%%%%%%%%%%%%%%%%%%%%%%%%%%%%%%%%%%%%%%%%
\section{Solving the SFP equation}
\label{ap:identities2}
We first give here several algebraic identities which will prove useful. From the definition of matrix $A = S^\dagger S$ we find
\begin{align}
\partial_{\bar{S}_{kl}} A_{\alpha m} & = S_{km} \delta_{l\alpha}, \qquad \partial_{S_{kl}} A_{\alpha \beta} = S^\dagger_{\alpha k}\delta_{l\beta}, \nonumber \\
\partial_{\bar{S}_{kl}} A^{-1}_{n\alpha} & = -A^{-1}_{nl} S^{\dagger,-1}_{k\alpha}, \qquad \partial_{S_{kl}} A^{-1}_{\alpha \beta} = - S^{-1}_{\alpha k} A^{-1}_{l\beta}. \label{derids}
\end{align}
\paragraph{Proof of Eq. \eqref{TQ}.} We first establish the identity:
\begin{align}
& T_Q = \sum_{i,j} C_{ij}^{\lambda,\bar{\lambda}} \partial_{\lambda_i \bar{\lambda}_j} F + \sum_{i,k,l}\partial_{\bar{S}_{kl}}C_{ikl}^{\lambda,\bar{S}} \partial_{\lambda_i}F + \sum_{i,k,l}\partial_{S_{kl}}C_{ikl}^{S,\bar{\lambda}} \partial_{\bar{\lambda}_i}F + \sum_{k,l, n,m} \partial_{S_{kl}, \bar{S}_{nm}} C_{klnm}^{S,\bar{S}} F = 0, \label{id11}
\end{align}
valid for $F(\Lambda) = \prod_{i<j} |\lambda_j - \lambda_i|^4$ and we denote $C_{ij}^{\lambda,\bar{\lambda}} = O_{ij}, \quad C_{klnm}^{S,\bar{S}} = \sum\limits_{\substack{\alpha (\neq l), \beta (\neq m)}} \frac{S_{k\alpha} \bar{S}_{n\beta} A_{ml} A^{-1}_{\alpha\beta}}{(\lambda_l - \lambda_\alpha)(\bar{\lambda}_m -\bar{\lambda}_\beta)}, 
C_{ikl}^{\lambda,\bar{S}} = \sum\limits_{n(\neq l)} \frac{\bar{S}_{kn} A_{li} A^{-1}_{in}}{\bar{\lambda}_l - \bar{\lambda}_n}$ and $C_{ikl}^{S,\bar{\lambda}} = \sum\limits_{n(\neq l)} \frac{S_{kn} A_{il} A^{-1}_{ni}}{\lambda_l - \lambda_n}$.

To this end, we first compute
\begin{align*}
\partial_{\lambda_i} F & = 2F\sum_{k (\neq i)} \frac{1}{\lambda_i - \lambda_k},\\ %\quad \text{[CH1]} 
\partial_{\lambda_i \bar{\lambda}_j} F & = 4F\sum_{k (\neq i)} \sum_{l (\neq j)} \frac{1}{(\lambda_i - \lambda_k)(\bar{\lambda}_j - \bar{\lambda}_l)}, %\quad \text{[CH2]} 
\end{align*}
and 
\begin{align}
\sum_{kl} \partial_{\bar{S}_{kl}} C^{\lambda, \bar{S}}_{ikl} & = - 2 \sum_{n\neq l} \frac{O_{il}}{\bar{\lambda}_l - \bar{\lambda}_n}, \quad \to \quad \sum_{ikl} \partial_{\bar{S}_{kl}} C^{\lambda, \bar{S}}_{ikl} \partial_{\lambda_i} F = - 4F \sum_{k\neq i} \sum_{n\neq l} \frac{O_{il}}{(\lambda_i - \lambda_k)(\bar{\lambda}_l - \bar{\lambda}_n)}, \\
\sum_{kl} \partial_{S_{kl}} C^{S, \bar{\lambda}}_{ikl} & = - 2 \sum_{n\neq l} \frac{O_{li}}{\lambda_l - \lambda_n} \quad \to \quad \sum_i \sum_{kl} \partial_{S_{kl}} C^{S, \bar{\lambda}}_{ikl} \partial_{\bar{\lambda}_i} F = - 4F \sum_{k\neq i} \sum_{n\neq l} \frac{O_{li}}{(\lambda_l - \lambda_n)(\bar{\lambda}_i - \bar{\lambda}_k)}, \label{result2}\\
F \sum_{klnm} \partial_{S_{kl}, \bar{S}_{nm}} C^{S, \bar{S}}_{klnm} & = 4F \sum_{\alpha \neq l, \beta \neq m} \frac{O_{\alpha \beta}}{(\lambda_l - \lambda_\alpha)(\bar{\lambda}_m - \bar{\lambda}_\beta)}, \\
\sum_{i,j} C_{ij}^{\lambda,\bar{\lambda}} \partial_{\lambda_i \bar{\lambda}_j} F & = 4F \sum_{k \neq i} \sum_{l \neq j} \frac{O_{ij}}{(\lambda_k - \lambda_i)(\bar{\lambda}_l - \bar{\lambda}_j)} ,
\end{align}
so that after renaming the indices all four contributions to the formula \eqref{id11} add up to zero. 

\paragraph{Proof of Eq. \eqref{TdQ}.} To prove the formula :
\begin{align*}
T_{\partial Q} & =  \sum_{j} \partial_{\bar{\lambda}_j} Q_t \left [ \sum_i C_{ij}^{\lambda,\bar{\lambda}} \partial_{\lambda_i } F + F \sum_{k,l}\partial_{S_{kl}}C_{jkl}^{S,\bar{\lambda}} \right ]  + \sum_{i}  \partial_{\lambda_i } Q_t \left [ \sum_j C_{ij}^{\lambda,\bar{\lambda}} \partial_{\bar{\lambda}_j} F + F\sum_{k,l}\partial_{\bar{S}_{kl}}C_{ikl}^{\lambda,\bar{S}} \right ] + \\
& + \sum_{k,l} \partial_{S_{kl}} Q_t \left [ \sum_i C_{ikl}^{S,\bar{\lambda}} \partial_{\bar{\lambda}_i} F +  F \sum_{n,m} \partial_{\bar{S}_{nm}} C_{klnm}^{S,\bar{S}} \right ] + \sum_{k,l} \partial_{\bar{S}_{kl}} Q_t \left [ \sum_i C_{ikl}^{\lambda,\bar{S}} \partial_{\lambda_i} F + F \sum_{nm} \partial_{S_{nm}} C_{nmkl}^{S,\bar{S}}\right ] = 0,
\end{align*}
we focus on showing that the terms in the brackets vanish:
\begin{align}
& \sum_i C_{ij}^{\lambda,\bar{\lambda}} \partial_{\lambda_i } F + F \sum_{k,l}\partial_{S_{kl}}C_{jkl}^{S,\bar{\lambda}} = 0, \label{id22}\\
& \sum_i C_{ikl}^{S,\bar{\lambda}} \partial_{\bar{\lambda}_i} F +  F \sum_{n,m} \partial_{\bar{S}_{nm}} C_{klnm}^{S,\bar{S}} = 0. \label{id4} 
\end{align}
The first identity \eqref{id22} is established by computing the first term
\begin{align*}
\sum_i C_{ij}^{\lambda,\bar{\lambda}} \partial_{\lambda_i } F & = 2F\sum_{k \neq i} \frac{O_{ij}}{\lambda_i - \lambda_k}
\end{align*}
and noticing that after renaming indices $i \to l, k \to n$ and $j \to i$ it cancels the second term calculated in Eq. \eqref{result2}. The second identity \eqref{id4} is found by computing
\begin{align}
F\sum_{nm} \frac{\partial}{\partial \bar{S}_{nm}} C^{S, \bar{S}}_{klnm} & = 2F \sum_{\alpha(\neq l)} \sum_{\beta \neq m} \frac{S_{k\alpha} A_{\beta l} A^{-1}_{\alpha\beta}}{(\lambda_l - \lambda_\alpha)(\bar{\lambda}_m - \bar{\lambda}_\beta)}, \\
\sum_i C_{ikl}^{S,\bar{\lambda}} \partial_{\bar{\lambda}_i} F & = 2F \sum\limits_{n(\neq l)} \sum_{k \neq i} \frac{S_{kn} A_{il} A^{-1}_{ni}}{(\lambda_l - \lambda_n)(\bar{\lambda}_i - \bar{\lambda}_k)}, \label{second}
\end{align}
which cancel each other as can be seen by renaming $n \to \alpha, i \to \beta$ and $k \to m$ in the formula \eqref{second}.

\section{Alternative integral representation of Eq. \eqref{Ispiric}}
\label{reprspiric}
We start off from Eq. \eqref{Ispiric}:
\begin{align*}
I_{\text{spiric}} & = \int_0^1 \dd \beta ~ \left [ \oint_{\mathcal{C}} \frac{\dd v}{2\pi i} \frac{e^{-\beta \frac{N}{\tau} v}}{\beta} \left ( 1 - \frac{\beta v}{v- A_-} \right )^{N/2} \left ( 1 - \frac{\beta v}{v- A_+} \right )^{N/2} \right ].
 \end{align*} 
First we expand both binomials:
\begin{align*}
I_{\text{spiric}} = \int_0^1 \dd  \beta \oint \frac{\dd v}{2\pi i} \frac{1}{\beta} \sum_{k,l=0}^{N/2}  \binom{N/2}{k} \binom{N/2}{l} \frac{(-\beta v)^{k+l} e^{-\beta \frac{N}{\tau} v}}{(v-A_-)^l(v-A_+)^k} %\quad \text{[CH12]}
\end{align*}
and split sums into $k,l=0$ and $k,l\neq 0$ indices resulting in four terms. The $k=l=0$ term vanishes upon taking the $v$ contour integral (and also yields the spurious $\beta$ integral). The three remaining are:
\begin{align*}
I_{\text{spiric}} = I_{A_+}^{(1)} + I_{A_-}^{(1)} + I_A^{(2)} , %\quad \text{[CH13]}
\end{align*}
where
\begin{align*}
I_{A_\pm}^{(1)} & = \sum_{k=1}^{N/2} \binom{N/2}{k}\frac{(-1)^k}{(k-1)!} \frac{d^{k-1}}{dv^{k-1}} \left [ v^k \left ( \int_0^1 \dd \beta \beta^{k-1} e^{-\beta \frac{N}{\tau} v} \right ) \right ]_{v = a_{\pm}}, \\
I_A^{(2)} & = \sum_{k=1}^{N/2} \sum_{l=1}^{N/2} \binom{N/2}{l} \binom{N/2}{k} \oint \frac{dv}{2\pi i} \int_0^1 \frac{\dd \beta }{\beta} \frac{(-\beta v)^{k+l} e^{-\beta N v/\tau}}{(v-a_-)^l(v-a_+)^k}.
\end{align*}
At this point all $\beta$ integrals are convergent. The first two terms have a structure identical to \eqref{I} and so we readily find:
\begin{align*}
I_{A_\pm}^{(1)} = - \frac{\tau}{2} + \frac{\tau}{N} e^{- \frac{N A_\pm}{\tau}} \sum_{m=0}^{N/2-1} \frac{N/2-m}{m!} \left (\frac{N A_\pm}{\tau} \right )^m, %\quad \text{[CH14]}
\end{align*}
or with the integral representation \eqref{Ivarintegr}:
\begin{align}
\label{I1pm}
I_{A_\pm}^{(1)} = - \frac{\tau}{2} - (-1)^{N/2} \tau e^{- \frac{N A_\pm}{\tau}} \int_0^\infty \frac{\dd u}{2\pi i} \oint_{C(0)} \dd \sigma e^{-Nu - N \sigma} \frac{1}{A_\pm/\tau + u + \sigma}\left ( \frac{A_\pm/\tau + u}{\sigma} \right )^{N/2}. %\quad \text{[CH35]}
\end{align}
The term $I_A^{(2)}$ is less trivial, we first compute the $\beta$ integral which gives
\begin{align*}
I_A^{(2)} =- \sum_{k=1}^{N/2} \sum_{l=1}^{N/2} \binom{N/2}{l} \binom{N/2}{k} \left ( -\frac{\tau}{N} \right )^{k+l} \oint \frac{\dd v}{2\pi i} \frac{\Gamma\left (k+l,\frac{N}{\tau} v \right )}{(v-A_-)^l(v-A_+)^k}, % \qquad \text{[CH15]}
\end{align*}
where we have used that $\sum_{k=1}^{N/2} \sum_{l=1}^{N/2} \binom{N/2}{l} \binom{N/2}{k} \left ( -\frac{\tau}{N} \right )^{k+l} (k+l-1)! \oint \frac{\dd v}{2\pi i} \frac{1}{(v-A_-)^l(v-A_+)^k} = 0$ based on the Cauchy theorem. The contour integral is explicitly evaluated:
\begin{align*}
I_A^{(2)}  = \sum_{k,l=1}^{N/2} \sum_{n=0}^{l-1} \sum_{m=0}^{l-n-1} \binom{N/2}{k} \binom{N/2}{l} \binom{k+l-1}{n} \frac{1}{(l-n-1)!} \binom{l-n-1}{m} \frac{(k+m-1)!}{(k-1)!} (-1)^{k-n} \left ( \frac{\tau}{N} \right )^{k+m+1} \times \nonumber \\
 \times \frac{1}{(A_- - A_+)^{m+k}} \left [ \Gamma \left (k+l-n,\frac{N}{\tau} A_- \right ) + (-1)^{m+k} \Gamma \left (k+l-n,\frac{N}{\tau} A_+ \right ) \right ]. %\quad \text{[CH16]}
\end{align*}
We decompose $I_A^{(2)} = I_{A_-}^{(2)} + I_{A_+}^{(2)}$:
\begin{align*}
I_{A_\pm}^{(2)}  = \sum_{k,l=1}^{N/2} \sum_{n=0}^{l-1} \sum_{m=0}^{l-n-1} & \binom{N/2}{k} \binom{N/2}{l} \binom{k+l-1}{n} \frac{1}{(l-n-1)!} \binom{l-n-1}{m} \frac{(k+m-1)!}{(k-1)!} (-1)^{k-n} \left ( \frac{\tau}{N} \right )^{k+m+1} \times \nonumber \\
 & \times \frac{1}{(A_\pm - A_\mp)^{m+k}} \Gamma \left (k+l-n,\frac{N}{\tau} A_\pm \right ).
\end{align*}
All these summations are expressible in terms of the confluent hypergeometric functions and in turn are rewritten in terms of integral representations $U(a,b,z) = \Gamma(a)^{-1} \int_0^\infty \dd t e^{-zt} t^{a-1} (1-t)^{b-a-1}$ valid for $a>0$ and $U(a,b,z) = e^{-a\pi i} \frac{\Gamma(1-a)}{2\pi i} \oint_{C(0)} \dd t e^{-zt} t^{a-1} (1+t)^{b-a-1}$ for $a\leq 0$. For $A_- > A_+$ we find:
\begin{align*}
I_{A_\pm}^{(2)} & = \pm \frac{\tau}{N} e^{-\frac{N}{\tau} A_\pm} \int_0^\infty \dd u \int_0^\infty  \frac{\dd t}{2\pi i} \oint_{C(0)}   \frac{\dd \sigma}{2\pi i} \oint_{C(0)} \dd \alpha \frac{1}{1 \mp t} \left ( \frac{(1-\alpha)(1 \mp t)\left (\frac{N}{\tau} A_\pm + u \right )}{\sigma \alpha} \right )^{N/2} \times \nonumber \\
& \times \exp \left ( -\frac{N}{\tau}(A_- -A_+)t - u - \sigma \mp \alpha t \left ( \frac{N}{\tau} A_\pm + u + \frac{\sigma}{1 \mp t} \right ) \right ). %\quad \text{[CH32]}
\end{align*}
We rescale $u \to Nu$ and $\sigma \to N\sigma(1 \mp t)$:
\begin{align*}
I_{A_\pm}^{(2)} & = \pm \tau N e^{-\frac{N}{\tau} A_\pm} \int_0^\infty \dd u \int_0^\infty  \frac{\dd t}{2\pi i} \oint_{C(0)}   \frac{\dd \sigma}{2\pi i} \oint_{C(0)} \dd \alpha e^{N f_\pm(u,t,\sigma,\alpha)},
\end{align*}
where $f_\pm(u,t,\sigma,\alpha)  = \frac{1}{2} \log \left [ \frac{(1-\alpha)\left (\frac{A_\pm}{\tau} + u \right )}{\sigma \alpha} \right ] - \frac{t}{\tau} (A_- - A_+) - u - (1\mp t)\sigma \mp \alpha t \left ( \frac{A_\pm}{\tau} + u + \sigma \right )$.
We first calculate the $t$ integral:
\begin{align*}
I_{A_\pm}^{(2)} = - \tau e^{-\frac{N}{\tau} a_\pm} \int_0^\infty  \frac{\dd u}{2\pi i} \oint_{C(0)} \dd \sigma \frac{1}{2\pi i} \oint_{C(0)} \dd \alpha  \frac{1}{A_\pm/\tau + u + \sigma}  \frac{1}{\alpha - \alpha^\pm_0} e^{-N\sigma - Nu} \left ( \frac{A_-/\tau + u}{\sigma} \right )^{N/2} \left ( \frac{1-\alpha}{\alpha} \right )^{N/2}, %\quad \text{[CH33]}
\end{align*}
where $\alpha^\pm_0 = \pm \frac{\frac{1}{\tau}(A_- -A_+) \pm \sigma}{A_\pm/\tau + u + \sigma}$ and we continue in calculating the contour integral over $\alpha$ variable. This is achieved by the Cauchy theorem -- instead of encircling pole of order $N$ at zero, we compute a simple pole at $\alpha_0^\pm$:
\begin{align}
\label{I2pm}
I_{A_\pm}^{(2)} & = -\tau e^{-\frac{N}{\tau} a_\pm} \int_0^\infty  \frac{\dd u}{2\pi i} \oint_{C(0)} \dd \sigma  \frac{(A_\pm/\tau + u)^{N/2}}{(A_\pm/\tau + u + \sigma)\sigma^{N/2}} \left [ \left ( \frac{A_\mp/\tau+u}{\sigma \mp \frac{A_- - A_+}{\tau} }\right )^{N/2} - (-1)^{N/2} \right ] e^{-Nu -N\sigma}. %\quad \text{[CH33]}
\end{align}
We collect formulas \eqref{I1pm} and \eqref{I2pm} to find
\begin{align*}
I_{A_\pm} = I_{A_\pm}^{(1)} + I_{A_\pm}^{(2)} = -\tau -\tau e^{-\frac{N}{\tau} a_\pm} \int_0^\infty  \frac{\dd u}{2\pi i} \oint_{C(0)} \dd \sigma  \frac{(A_\pm/\tau + u)^{N/2}}{(A_\pm/\tau + u + \sigma)\sigma^{N/2}} \left ( \frac{A_\mp/\tau+u}{\sigma \mp \frac{A_- - A_+}{\tau} }\right )^{N/2}  e^{-Nu -N\sigma}.
\end{align*}
We lastly deform $\sigma \to \frac{1}{\tau} (\sigma - A_\pm), u \to \frac{u}{\tau}$ and finally obtain 
\begin{align*}
I_{A_\pm} = - \tau - \int_0^\infty  \frac{\dd u}{2\pi i} \oint_{C(A_\pm)} \dd \sigma \frac{e^{-\frac{N}{\tau} (u +\sigma)}}{u+\sigma} \left ( \frac{(A_+ + u)(A_- + u)}{(\sigma-A_-)(\sigma-A_+)} \right )^{N/2}.
\end{align*}
Summing these two contributions, $I_{spiric}$ is rewritten as a single contour integral 
\begin{align*}
I_{spiric} & = - \tau - \int_0^\infty \frac{\dd u }{2\pi i} \oint_{C(A_-,A_+)} \dd \sigma \frac{e^{-\frac{N}{\tau} (u +\sigma)}}{u+\sigma} \left ( \frac{(A_+ + u)(A_- + u)}{(\sigma-A_-)(\sigma-A_+)} \right )^{N/2}. %\qquad \text{[CH36]}
\end{align*}
The formula does not depend on which $A_\pm$ is larger and so we recreate Eq. \eqref{Ispiric2} valid for any $A_\pm$.
%%%%%%%%%%%%%%%%%%%%%%%%%%%%%%%%%%%%%%%%%%%%%%%%%%%%%%%%%%%%%%%%%%%%%%
%%%%%%%%%%%%%%%%%%%%%%%%%%%%%%%%%%%%%%%%%%%%%%%%%%%%%%%%%%%%%%%%%%%%%%
%%%%%%%%%%%%%%%%%%%%%%%%%%%%%%%%%%%%%%%%%%%%%%%%%%%%%%%%%%%%%%%%%%%%%%
%%%%%%%%%%%%%%%%%%%%%%%%%%%%%%%%%%%%%%%%%%%%%%%%%%%%%%%%%%%%%%%%%%%%%%

%%%%%%%%%%%%%%%%%%%%%%%%%%%%%%%%%%%%%%%%%%%%%%%%%%%%%%%%%%%%%%%%%%%%%%
%%%%%%%%%%%%%%%%%%%%%%%%%%%%%%%%%%%%%%%%%%%%%%%%%%%%%%%%%%%%%%%%%%%%%%
%%%%%%%%%%%%%%%%%%%%%%%%%%%%%%%%%%%%%%%%%%%%%%%%%%%%%%%%%%%%%%%%%%%%%%

\end{document}